\documentclass[prx,longbibliography,superscriptaddress,twocolumn,amsmath,amssymb]{revtex4-1}

\usepackage{amsmath,amssymb}
\usepackage{tabularx}
\usepackage[dvipdfmx]{graphicx}
\usepackage{bmpsize}
\usepackage{bm}
\usepackage{color}
\DeclareMathAlphabet{\mathpzc}{OT1}{pzc}{m}{it}
\newcommand{\nn}{\nonumber}

\begin{document}

\preprint{APS/123-QED}

\title{Topological crystalline materials of $J=3/2$ electrons:  \\
antiperovskites, Dirac points, and {high winding} topological superconductivity} 

\author{Takuto Kawakami}
\email{takuto.kawakami@yukawa.kyoto-u.ac.jp}
\affiliation{Yukawa Institute for Theoretical Physics, Kyoto University, Kyoto 606-8502, Japan}
\author{Tetsuya Okamura}
\affiliation{Yukawa Institute for Theoretical Physics, Kyoto University, Kyoto 606-8502, Japan}
\author{Shingo Kobayashi}
\affiliation{Institute for Advanced Research, Nagoya University, Nagoya 464-8601, Japan}
\affiliation{Department of Applied Physics, Nagoya University, Nagoya 464-8603, Japan}
\author{Masatoshi Sato}
\email{msato@yukawa.kyoto-u.ac.jp}
\affiliation{Yukawa Institute for Theoretical Physics, Kyoto University, Kyoto 606-8502, Japan}

\date{\today}

\begin{abstract} 
We present a theory of the high-spin generalization of topological insulators and their doped superconducting states.
The higher-spin topological insulators involve a pair of $J=3/2$ bands with opposite parity, 
and are characterized  by their band inversion.
The low-energy effective theory reveals that the topological insulators host four different phases characterized by 
mirror Chern numbers, at which boundaries two different patterns of bulk Dirac points appear.
For the carrier-doped case,
it is shown that the {system} may host unique unconventional superconductivity 
because of its high-spin nature and additional orbital degrees of freedom intrinsic to topological insulators. 
The superconducting critical temperature is evaluated by
using density-density pairing interactions, and odd-parity Cooper pairs are shown to be naturally realized {in the presence of interorbital pairing interaction}. 
It is observed that {even} the simplest spin 0 odd-parity pairing state exhibits 
{a novel class of topological superconductivity---high winding topological superconductivity}. 
{We also discuss the experimental signals of high winding topological superconductivity in the case of the antiperovskite superconductor Sr$_{3-x}$SnO.} 
\end{abstract}


\maketitle
\section{Introduction}
The search for topological materials is a recent trend in
condensed matter physics~\cite{hasan2010, qi2011, TSN2012, Andoreview, Ando-Fu2015, mizushimareview2016, Sato-Fujimoto2016, sato2017}. 
A promising direction to fulfil this search is a multi-orbital
material with strong spin-orbit coupling.
Such a system can be topologically non-trivial through band inversion owing to
spin-orbit coupling. 
For instance, in the case of topological
insulators (TIs), the $\bm{Z}_2$ topological indices 
are directly related to the number of time-reversal
invariant momenta at which band inversion occurs~\cite{fu2007,qi2008}.
Following this insight, 
numerous TIs including Bi$_2$Se$_3$ and Bi$_2$Te$_3$ have been
discovered experimentally~\cite{hasan2010,Andoreview}.

In the case of superconductors, 
even a single orbital system may possess a 
non-trivial topological phase because the charge conjugate counterpart of the single band coexists in the Bogoliubov--de Gennes (BdG)  Hamiltonian~\cite{VolovikHe3, Goryo1999,
read2000,schnyder2008, Qi2009, Roy2008, Sato2009}.  
Any spin-triplet superconductor can be topological if the Fermi surface is properly chosen~\cite{Sato2009}. 
However, the search for spin-triplet superconductivity itself is a challenge
because of its strongly correlated origin.
No spin-triplet superconductor has been established yet, despite the existence of several promising candidates. 

Recently, it has been recognized that 
multi-orbital systems may solve this difficulty, as they
allow another mechanism of topological superconductivity~\cite{sato2017}:
Using interorbital pairing interaction and spin-orbit coupling, 
a multi-orbital system may host
odd-parity Cooper pairs even without strong correlation~\cite{fu2010},
which indicates topological superconductivity~\cite{Sato2009, Sato2010, fu2010}.
In this new direction,
various types of topological superconductivity  in doped topological materials
have been discussed recently~\cite{fu2010,sasaki2011, Fu2014, Matano2016, Yonezawa2016, Liu2015, Qiu2015, Maruya2015, Asaba2017,
Murakami-Nagaosa2003,Cho2012, Bo2015, Li-Haldane2015, aggrwal2015, wang2015,
Benik2015, kobayashi2015, hashimoto2016, shen2017}.

In this study, we explore another class of topological
phases that can be realized only in $J=3/2$ bands.
Spin-orbit coupled electrons may behave as {higher-order spin states}, 
owing to the mixture of spin and orbital angular momentum. 
In crystals, the higher spin state can be identified as a $J=3/2$ state, 
as discrete crystalline rotation allows at most fourfold degeneracy of the $J=3/2$ spin.
Superconductivity of $J=3/2$ electrons
has been discussed for Luttinger semimetals with a single $J=3/2$ band such as half-Heusler systems~\cite{brydon2016,yang2016, boettcher2016, savary2017, boettcher2018, roy2017, timm2017, ghorashi2017, venderbos2018, ho1999, kuzmenko2018, yu2018}.
While the half-Heuslers may host interesting higher-spin Cooper pairs, 
their actual realization is restricted because the systems support
only a single $J=3/2$ band.

Here, alternatively, we consider systems with multiple $J=3/2$ bands.
A class of relevant materials is antiperovskite $A_3BX$, where $A$ is (Ca, Sr, La), 
$B$ is (Pb, Sn), and $X$ is (C, N, O). 
The antiperovskite materials support two different $J=3/2$
electrons near the Fermi level, i.e., $d$-orbital electrons of the $A$ atom and
$p$-orbital ones of the $B$.
The first-principle calculations show that the band inversion of these
two orbitals may occur at the $\Gamma$-point, accompanied with
three-dimensional (3D) Dirac points with a tiny gap~\cite{kariyado2011,
kariyado2012, hsieh2014}.
Owing to the band inversion, these materials become topological
crystalline insulators~\cite{hsieh2014, fang2017}.
It was also discovered very recently that
one of the antiperovskite topological materials, Sr$_3$SnO, shows
superconductivity with hole doping~\cite{oudah2016, hausmann2017, ikeda2018, oudah2018}.
Although discussion from the
viewpoint of $J=3/2$ electrons was missing, 
possible topological
superconductivity in Sr$_{3-x}$SnO was suggested
theoretically~\cite{oudah2016}, {using} the analogy of
the superconducting Dirac semimetal, Cd$_3$As$_2$~\cite{kobayashi2015, hashimoto2016}.

Herein, we develop a general theory of higher-spin topological materials. 
Considering the application to antiperovskite,
we assume cubic symmetry, which allows for four-dimensional representation corresponding to the fourfold degeneracy of $J=3/2$.
Based on the low-energy effective Hamiltonian, we observe that the system hosts four different topological crystalline insulating phases, where the phase boundaries are characterized by two different patterns of gapless Dirac points. 
In addition to the octahedral Dirac points discussed previously~\cite{kariyado2011,
kariyado2012, hsieh2014},
cubic Dirac points appear at one of the phase boundaries.  

We thereafter study the superconducting states of doped higher-spin topological materials.   
We first classify possible momentum-independent
gap functions, based on the cubic symmetry.
The gap functions may contain spin $J=2$
(spin-quintet) and $J=3$ (spin-septet) components, and ordinary $J=0$ (spin-singlet) and $J=1$ (spin-triplet) ones,
as Cooper pairs are formed by the $J=3/2$ electrons.
In contrast to the single $J=3/2$ band case as in half-Heusler materials, 
additional orbital degrees of freedom make it possible to obtain any higher-spin gap function 
{in the framework of weakly correlated constant gap functions.}
Subsequently, we evaluate the superconducting critical temperature $T_{\rm c}$ for each gap
function by using simple density-density pairing interactions. 
It is observed that the system supports odd-parity superconductivity if the
interorbital pairing interaction is dominant.
The $J=0$ odd-parity pairing state with
$A_{1u}$ representation yields the highest $T_{\mathrm{c}}$ among odd-parity superconducting states,
but the $T_{1u}$ paring state consisting of $J=3$ and $J=1$ Cooper pairs also has high $T_{\rm c}$.

We reveal that the simplest $J=0$ odd-parity pairing state shows a new class of topological superconductivity
i.e., high winding topological superconductivity. 
As a 3D time-reversal invariant superconductor,
the odd-parity pairing state has a non-zero 3D winding number similar to the $^3$He-B phase. 
However, in contrast to $^3$He-B, owing to the higher spin nature of $J=3/2$ electrons, 
it does not support the minimal non-zero value but hosts various higher winding numbers 
depending on the model parameters. 
It is predicted that the topological superconductivity with a higher winding number 
exhibits a higher value of quantized longitudinal thermal conductance, 
which is proportional to the winding number~\cite{foster2014,roy2017}. 
Moreover, the accompanying multiple surface Majorana fermions 
display a new physics of conformal field theory with non-Abelian gauge field~\cite{foster2014}. 
Furthermore, owing to a non-trivial spin texture on the Fermi surface 
originating from the winding number, we can expect non-trivial magnetic responses.
We also identify the topological phase diagram of the superconducting state 
and determine the relevant topological numbers and the patterns of surface Majorana fermions. 
We finally discuss the experimental signals of high winding superconductivity in the case of Sr$_{3-x}$SnO. 
We observe that Sr$_{3-x}$SnO is in the vicinity of topological phase transition, 
which predicts the characteristic nodal structure of the superconducting gap 
if it realizes the $J=0$ odd-parity pairing state.

This paper is organized as follows. 
In Sec.~\ref{sec:normal}, we examine the normal state of higher-spin topological materials. 
Using the general $k\cdot p$ Hamiltonian involving two $J=3/2$ bands with opposite parity, we obtain the phase diagram of the $J=3/2$ topological crystalline insulators.
We reproduce the results for the antiperovskite oxides~\cite{kariyado2011, kariyado2012, hsieh2014}, clarifying {the} topological origin of Dirac points in antiperovskites. 
We also observe that a novel cubic pattern of Dirac points is possible, in
addition to the octahedral Dirac points in antiperovskites. 
In Sec.~\ref{sec:super}, we examine the superconductivity of doped higher-spin TIs.
We classify gap functions 
in terms of an irreducible representation of the $O_h$ point group, 
and evaluate the superconducting critical temperature
for each gap function. 
We demonstrate that the $A_{1u}$ ($J=0$) representation is the most stable odd-parity pairing state, 
and the $T_{1u}$ ($J=3$, or $1$) is the second one. 
In Sec.~\ref{sec:A1u}, we identify the topological crystalline superconductivity of the $A_{1u}$ state. 
Sec.~\ref{sec:discussions} discusses the application of our theory 
to Sr$_{3-x}$SnO, the case without band inversion, and the property of other pairing states, 
and subsequently, the conclusion is presented in Sec.~\ref{sec:conclusions}.

\section{Normal state}\label{sec:normal}
\subsection{Effective Hamiltonian}

Here, we formulate higher-spin TIs. 
First, we define high-spin fermions. In the presence of spin-orbit coupling, 
electrons may behave as spin $J$ fermions with $J=|l\pm 1/2|$ owing to 
a mixture of the orbital angular momentum $l$ and the electron spin $1/2$. 
For electrons in crystals, however, a more precise definition of $J$ is necessary, 
as crystals partially break rotation symmetry. Considering that spin $J$ in the above sense  
indicates $(2J+1)$-fold degeneracy at $\bm{k}=0$, we define spin $J$ in crystals as a
$(2J+1)$-dimensional representation of the point group at the $\Gamma$ point.
From this definition, it is observed that 
only a $J=3/2$ fermion is possible as a
higher-spin fermion, as point groups allow a maximum of four-
dimensional representations at $\Gamma$.
Furthermore, cubic symmetry is necessary to obtain the four-dimensional
representation.
Therefore, we consider a cubic crystal in the following analysis.
We also assume inversion symmetry for simplicity, which specifies the
cubic symmetry as the $O_h$ group.

For ordinary spin $1/2$ {electrons}, a topological insulating phase 
with inversion symmetry is
obtained via band inversion between the orbitals with different parities at
the $\Gamma$ point. 
We can consider a similar band inversion mechanism even for $J=3/2$
electrons.
The $O_h$ group hosts two different four-dimensional, i.e., $J=3/2$,
representations, $G_{3/2g}$ and $G_{3/2u}$, corresponding to the even- and odd-parity states, respectively.
When these bands are inverted, the gap between
them closes at $\Gamma$, and thus, we can expect a non-trivial
topological phase {transition} through the band inversion.

This argument specifies a system for a higher-
spin TI.
It is symmetric under the $O_h$ group, and consists of two
different 
$J=3/2$ bands {corresponding} to 
the $G_{3/2g}$ and $G_{3/2u}$ representations of the $O_h$ group.  
Moreover, time-reversal symmetry should be assumed when we consider an
insulator.  

The single-particle state of the $J=3/2$ TI is
represented by an eight-
component spinor ${\bm c}_{\bm k}$ composed of the annihilation operator $c_{j_z, \sigma_z, \bm{k}}$, 
where $j_z=\pm 3/2, \pm 1/2$ denotes the $z$-component of $J$
corresponding to the four-dimensional representation of the $O_h$ group,
$\sigma_z$ specifies the parity of the representation, i.e., $\sigma_z=1$
($\sigma_z=-1$) for the $G_{3/2g}$ ($G_{3/2u}$) representation with even (odd) parity, 
which can be characterized as $d$, $g$, ($p$, $f$), or higher-order orbitals, 
and ${\bm k}$ is the momentum.
Generators of the $O_h$ group 
consist of $q$-fold discrete crystal rotations $C_{q,{\bm n}}$ ($q=2,3,4$) and inversion $P$, which act on the Hamiltonian $H_0({\bm k})$ as
\begin{eqnarray}
	&C_{q,\bm{n}}H_0(\bm{k}) C_{q,\bm{n}}^{-1}\!=\! H_0(D_{C_{q,\bm{n}}}[\bm{k}]) \ \hbox{with}\ C_{q,\bm{n}}\!=\!e^{-i\frac{2\pi}{q}\bm{J}\cdot\bm{n}},& \label{eq:Cqn}\\
	&PH_0(\bm{k}) P^{-1} = H_0(-\bm{k}) \ \hbox{with}\ P = \sigma_z. \label{eq:inv}&
\end{eqnarray}
Here, $\bm{n}$ denotes the rotation axis, $D_{C_{q,\bm{n}}}[\bm{k}]$ indicates the $q$-fold rotation of $\bm{k}$, $J_{i=x,y,z}$ is the $4\times 4$ $J=3/2$ spin operator (see Appendix~\ref{sec:Jmat}), and $\sigma_i$ is the 2$\times$2 Pauli matrix in the orbital space.
As the $G_{3/2u}$ ($G_{3/2g}$) state changes (does not change) the sign under inversion, 
$P$ is given as $P=\sigma_z$.
We also have time-reversal symmetry, 
\begin{eqnarray}\label{eq:TRS_N}
	&\mathcal{T}H_0(\bm{k}) \mathcal{T}^{-1} = H_0(-\bm{k}) \ \hbox{with}\ \mathcal{T} =  C_{2,\bm{y}}\mathcal{K}. &
\end{eqnarray}
with the complex conjugate operator $\mathcal{K}$.

In the following analysis, we use the low-energy effective Hamiltonian
for the $J=3/2$ electrons.  
To construct the effective Hamiltonian, we consider possible scalar matrices with respect to $O_h$ symmetry.
First, we note that $\sigma_z$ and $\sigma_0$ are scalars, whereas $\sigma_x$ and $\sigma_y$ are pseudo 
scalars.  
Any $\sigma_\nu$ ($\nu=0,x,y,z$) is invariant under rotation, but $\sigma_x$ and $\sigma_y$ change their signs under inversion. 
Second, as the spin operators $J_i$ $(i=x,y,z)$ 
and their third-order polynomials $\tilde{J}_i\equiv\frac{5}{3}\sum_{j\neq i}J_jJ_iJ_j-\frac{7}{6}J_i$ 
(see also Appendix~\ref{sec:Jmat})
behave as pseudo vectors for the $O_h$ point group,
the inner product $\bm{k}\cdot\bm{J}$ and $\bm{k}\cdot\tilde{\bm{J}}$ are pseudo scalars~\cite{elder2011}.
Finally, we can construct a scalar as a product of any two of the above pseudo
scalars. 
Consequently, the effective Hamiltonian for the lower order of momentum
$\bm{k}$ is given as~\cite{hsieh2014}
\begin{eqnarray}\label{eq:normal}
	H_0(\bm{k})	= m(\bm{k}) \sigma_z + \bm{k}\cdot (v_1\bm{J} + v_2\tilde{\bm{J}}) \sigma_x - \mu \sigma_0,
\end{eqnarray}
where the $\sigma_y$ term is absent owing to the time-reversal symmetry.

\begin{figure}[t]
\includegraphics[width=85mm]{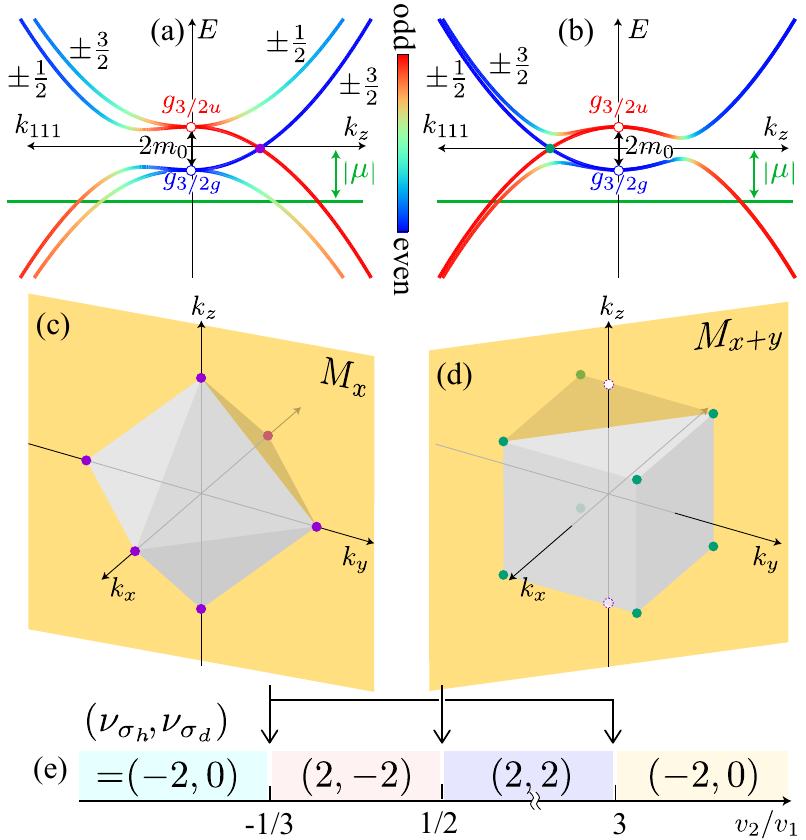}
\caption{ Energy dispersion near the $\Gamma$ point for (a) $v_2=3v_1$ and (b) $v_2=v_1/2$. 
The dispersions along $k_z$ and $k_{111}=(k_x+k_y+k_z)/\sqrt{3}$ are labeled by the $z$- and $(111)$-component of angular momenta $j_z$ and $j_{111}$, respectively.  
For the hole-doped antiperovskite material Sr$_{3-x}$SnO, the Fermi level is located below the fourfold degeneracy point $E=-m_0$ as indicated by the green line. (c) and (d) show the octahedral and cubic Dirac points for $v_2=3v_1$ and $v_2=v_1/2$, respectively. 
The dots denote the Dirac points. The transparent panels indicate the mirror plane defining the topological phase 
transition with the creation of the Dirac points. (e) shows the horizontal and diagonal mirror Chern numbers 
$\nu_{\sigma_{h}}$ and $\nu_{\sigma_{d}}$ as a function of $v_2/v_1$.
}
\label{fig:Ndirac}
\end{figure}

The effective mass $m({\bm k})$ in Eq.~(\ref{eq:normal}) is given as 
\begin{eqnarray}
	m(\bm{k}) = -m_0 + \alpha k^2 + \left[\bm{k}\cdot(\beta_1 \bm{J}\!+\!\beta_2 \tilde{\bm{J}})\right] \left[\bm{k}\cdot(\beta_3 \bm{J}\!+\!\beta_4\tilde{\bm{J}}) \right],\nn
\end{eqnarray}
up to the second order of $\bm{k}$, 
where $2m_0$ is the energy gap between the $G_{3/2u}$ and $G_{3/2g}$ states at the $\Gamma$
point.
In the following analysis, we focus on the case where band inversion
occurs at the $\Gamma$ point.
We consider $m_0>0$, $\alpha>0$, and $\beta_i=0$ $(i=1,2,3,4)$, which can
reproduce the band structure of Sr$_3$SnO qualitatively.
The case without band inversion will be briefly discussed in Sec.~\ref{sec:nobandinv}.

The second term of (\ref{eq:normal}) with $\sigma_x$ leads to
orbital mixing between even- and odd-parity states, which is represented in 
green color in Figs.~\ref{fig:Ndirac}(a) and (b). 
This mixture opens the global energy gap around $E=\mu$ 
except for the parameter points displayed in Fig~\ref{fig:Ndirac}.
Note that the $\tilde{\bm{J}}$ term reduces the continuous rotational symmetry to the discrete $O_h$ one.

The $\sigma_0$ term in Eq.~(\ref{eq:normal}) provides the Fermi level. 
In this section, to demonstrate the topological property of the band gap discussed above, 
we set $\mu=0$, where the Fermi level lies in the band gap and the system falls into the insulating state.
The large $|\mu|$ pushes the energy gap below or above the Fermi level. 
In this case, the system is a metal rather than an insulator, 
although the topological property of the band gap remains unchanged.
This metallic state is important to induce topological superconductivity, which will be discussed later.

Note that $H_0({\bm k})$ cannot be an ordinary TI, 
although band inversion can occur at the $\Gamma$ point.
The ${\bm Z}_2$ number for TI is always trivial owing to the fourfold
degeneracy of the $J=3/2$ spin at $\Gamma$.
However, the system can be a topological crystalline
insulator, as shown below.

\subsection{Octahedral Dirac points}

By diagonalizing the Hamiltonian~(\ref{eq:normal}), we obtain the band
spectrum.
First, let us focus on the band spectrum along the primary axis,
say the $k_z$-axis. 
On the $k_z$-axis, we have $[C_{4, \bm{z}}, H_0(0,0,k_z)]=0$ owing to the
fourfold rotation symmetry of the $O_h$ group,
$C_{4,\bm{z}}H_0(-k_y,k_x,k_z)C_{4,\bm{z}}^{-1}=H_0(\bm{k})$, and thus, 
the Hamiltonian is subdivided into the eigensector of $C_{4,\bm{z}}$ 
with the eigenvalue $\lambda_{C_{4,\bm{z}}}=e^{-i \frac{j_z}{2} \pi}$, 
\begin{eqnarray}\label{eq:mzsector}
	H_{0,j_z}(k_z) = m_{j_z}(k_z)\sigma_z + b_{j_z}k_z\sigma_x, 
\end{eqnarray}
where $m_{j_z}(k_z)=m_{-j_z}(k_z)$ is an even function of $k_z$ and $b_{j_z}$ is the
following constant:
\begin{eqnarray}
	b_{\pm 3/2} = \pm (3v_1-v_2)/2 \label{eq:bmz1},\label{eq:b32} \\
	b_{\pm 1/2} = \pm (v_1+3v_2)/2 \label{eq:bmz2}. \label{eq:b12}
\end{eqnarray}
From time-reversal and inversion symmetry, 
the sectors with opposite $j_z$ form a Kramers pair with the
same spectra.
The first term of Eq.~(\ref{eq:mzsector}) provides a parabolic energy
dispersion for each orbital, and the second one mixes them.
As shown in Eq.~(\ref{eq:b32}) (Eq.~(\ref{eq:b12})), the orbital mixing
in the $j_z=\pm 3/2$ ($j_z=\pm 1/2$) sector vanishes when
$v_2=3v_1$ ($v_1=-3v_2$). In such cases, Dirac points appear in
the band spectrum.

Figure~\ref{fig:Ndirac}(a) illustrates the band spectra
for $v_2=3v_1$.
The two parabolic dispersions 
with opposite orbital parity in the spin $j_z=\pm3/2$ sector
linearly cross each other at $k_z=\pm k_0 $ with $k_0$ satisfying
$m_{\pm 3/2}(k_0)=0$.  
Consequently, there appear two Dirac points on the $k_z$ axis. 
From the $O_h$ group symmetry, their counterparts also exist on the
$k_x$ and $k_y$ axes, respectively, forming 
vertices of an octahedron in the momentum space. See
Fig.~\ref{fig:Ndirac}(c).    
Further, 
orbitals with opposite parity in the $j_z=\pm1/2$ sectors hybridize with each
other and thus, they do not form a Dirac point.
Similarly, when $v_2=-v_1/3$, octahedral Dirac points appear in the
$j_z=\pm1/2$ sectors.

In contrast to ordinary Dirac semimetals such as Cd$_3$As$_2$, these Dirac
points are easily gapped out by small perturbation even if 
crystalline symmetry is preserved.
Thus, they are not stable, but their existence is not accidental.

They indeed appear as gap-closing points accompanied by a topological phase
transition.
A key topological number is the mirror Chern number.
Consider one of three equivalent horizontal mirror reflections of the
$O_h$ group, say the mirror
reflection with respect to the $yz$ plane:
\begin{eqnarray}
M_{x}H_0(-k_x,k_y,k_z)M_{x}^{-1}=H_0(\bm{k}),
\end{eqnarray} 
with $M_{x}=PC_{2,\bm{x}}$. See Fig.~\ref{fig:Ndirac} (c).    
On the mirror-invariant $k_x=0$ plane, it holds that $[M_{x},
H_0(0,k_y,k_z)]=0$, and hence, the Hamiltonian has the block diagonal form 
$H_0(0,k_y,k_z)=\mathrm{diag}[H_{0}^{+i}(0, k_y, k_z),H_{0}^{-i}(0, k_y, k_z)]$ 
in the eigenbasis of $M_x$ with the eigenvalues $\lambda_{M_x}=\pm i$.
Each sector may host its own Chern number

\begin{eqnarray}
	\nu_{\rm Ch}^{\pm i} = -i \sum_{\xi_{n}^{\pm i}<0}\int_{k_x=0} d^2k \epsilon^{\alpha\beta} \langle \partial_{\alpha} u_{n,\bm{k}}^{\pm i}| \partial_{\beta}u_{n,\bm{k}}^{\pm i} \rangle,
\end{eqnarray}
where we use the solution of the Bloch equation $H_{0}^{\pm i}|u_{n,\bm{k}}^{\pm i}\rangle=\xi_{n}^{\pm i}|u_{n,\bm{k}}^{\pm i}\rangle$  
and two-dimensional Levi-Civita symbol $\epsilon^{\alpha\beta}$ with $\alpha,\beta\in\{k_y,k_z\}$.
The mirror Chern number $\nu_{\sigma_h}$ for the horizontal mirror
reflection is defined by $\nu_{\sigma_h}=(\nu^{+i}_{\rm Ch}-\nu^{-i}_{\rm Ch})/2$.
From time-reversal symmetry, the total Chern number $\nu_{\rm
Ch}^{+i}+\nu_{\rm Ch}^{-i}$ vanishes, and thus, $\nu_{\sigma_h}$ is an
integer.

As summarized in Fig.~\ref{fig:Ndirac}(e), the mirror Chern number 
in the present system is
obtained as~\cite{hsieh2014}
\begin{eqnarray}\label{eq:chh0+}
	\nu_{\sigma_h}= 2\mathrm{sign}(b_{3/2}b_{1/2}),
\end{eqnarray}
which jumps by four when Dirac points appear at $v_2=3v_1$ {or}
$v_2=-v_1/3$.
In other words, each of the four Dirac points on the $k_x=0$ plane change
the mirror Chern number by one. 
Therefore, these Dirac points are caused by the topological phase
transition. 
This indicates that the Dirac points must appear in the phase diagram because they separate two distinct topological phases, 
even though for any specific choice of parameters (i.e., one point in the phase diagram) they can gapped.
The small perturbation, such as mixing from other bands not considered in the low-energy effective Hamiltonian~(\ref{eq:normal}), may open a gap in the Dirac points, but it
merely shifts the phase boundary of the topological phase transition,
which also hosts gapless Dirac points. 
Except at the topological phase transition, the system supports a
non-zero mirror Chern number, and thus,
the normal state is a topological crystalline insulator 
if the Fermi level is in the band gap.

\subsection{Cubic Dirac points}

The effective Hamiltonian (\ref{eq:normal}) also reveals that another class
of Dirac
points is possible in higher-spin topological materials.
To demonstrate this, consider the diagonal axis, say the $(111)$ axis. 
From the threefold rotational symmetry of the $O_h$ group, 
the Hamiltonian along the $(111)$ axis in momentum space 
is subdivided into eigensectors of $C_{3,(111)}$ 
with the eigenvalues $\lambda_{C_{3,(111)}}=e^{-i \frac{2}{3}j_{111}\pi}$,
where $(111)=(\hat{\bm x}+\hat{\bm y}+\hat{\bm z})/\sqrt{3}$ and $j_{111}=\pm 1/2, \pm 3/2$.
Note that $j_{111}=\pm 3/2$ falls into the same eigensector with $\lambda_{C_{3,(111)}}=-1$.
The Hamiltonian in the subsector takes the same form as 
Eq.~(\ref{eq:mzsector}), but $k_z$, $j_z$, and $b_{j_z}$ in Eq.~(\ref{eq:mzsector})
are replaced with $k_{111}=(k_x+k_y+k_z)/\sqrt{3}$, $j_{111}$, and $b_{j_{111}}'$, respectively.
The interorbital coupling $b_{j_{111}}'$ is given by
\begin{eqnarray}
	&&b_{\pm 1/2}' = \pm \frac{v_1-2v_2}{2}, \\
	&&b_{\pm 3/2}' = \left[ \left( \frac{3}{2}v_1 + \frac{1}{3}v_2 \right) \eta_z +  \frac{5\sqrt{2}}{6}v_2 \eta_x \right].
\end{eqnarray}
Here, $\eta_i$ is the Pauli matrix in the 
$j_{111}=\pm 3/2$ space.

{$b_{\pm 3/2}'$} does not vanish for any $v_i$, whereas {$b_{\pm 1/2}'$} becomes zero 
when $v_2=v_1/2$.   
In the latter case, the orbitals with opposite parity do not hybridize with each
other along
the $(111)$ axis, forming a Dirac point in
Fig.~\ref{fig:Ndirac}(b).   
By considering $O_h$ symmetry, we have a total of eight Dirac
points at the vertices of a cube, as illustrated in
Fig.~\ref{fig:Ndirac}(d).

The cubic Dirac points are also understood as gap-closing points for
the topological phase transition.  
The relevant topological index in the present case is the mirror Chern
number for the diagonal mirror reflection of the $O_h$ group.
For instance, consider the diagonal mirror reflection with respect to
the $k_x+k_y=0$ plane in Fig.~\ref{fig:Ndirac}(d).   
Similar to the octahedral Dirac case,
we can evaluate the mirror Chern number $\nu_{\sigma_d}$ with respect to
the diagonal
mirror plane. The result is shown in Fig.~\ref{fig:Ndirac}(f). 
At $v_2=v_1/2$, where there are four Dirac points on this plane,
the mirror Chern number jumps by four. 

Notably, two of the octahedral Dirac points appear on
the diagonal mirror plane when $v_2=3v_1$ or $v_2=-v_1/3$. 
Consequently, the diagonal mirror Chern number in Fig.~\ref{fig:Ndirac}(f)
jumps by two at $v_2=3v_1$
and $v_2=-v_1/3$.

\section{Superconductivity}
\label{sec:super}

\subsection{$O_h$ classification of the gap function}\label{sec:class}

With carrier doping, TIs can be superconductors at
low temperature. 
Here, we consider the superconducting states of doped higher-spin TIs.

To describe the superconducting states,
we use the Nambu space spanned by the basis 
$(\bm{c}_{\bm{k}},\ \bar{\bm{c}}_{-\bm{k}})$ 
with the eight-component spinor $\bm{c}_{\bm{k}}$ of the annihilation operator 
and its time-reversal hole partner $\bar{\bm c}_{-\bm k}$ with the component 
$\bar{c}_{j_z, \sigma_z, -\bm{k}}=\sum_{j_z'} (C_{2,\bm{y}})_{j_z,j_z'}c_{j_z', \sigma_z, -\bm{k}}^\dag$.
In this basis, the BdG Hamiltonian is written as 
\begin{eqnarray}\label{eq:bdg}
	H(\bm{k}) \!=\! H_0({\bm k})\tau_z + \mathrm{Re}(\Delta({\bm k}))\tau_x + \mathrm{Im}(\Delta({\bm k}))\tau_y
\end{eqnarray}
where $\tau_\mu$ represents the $2\times 2$ matrices acting on the particle and time-reversal hole space, 
and we have used time-reversal symmetry for the one-particle Hamiltonian 
$\mathcal{T}H_0({\bm k})\mathcal{T}^{-1}=H_0(-{\bm k})$. 
Note that, in this basis, the point group and time-reversal act on the gap
function $\Delta({\bm k})$ in the same manner as Eqs.~(\ref{eq:Cqn}), (\ref{eq:inv}), and (\ref{eq:TRS_N}) on the normal Hamiltonian
$H_0({\bm k})$
(see Appendix~\ref{sec:PTSC} for more details).

We classify the multi-component gap function $\Delta({\bm k})$.
For simplicity, we assume that $\Delta({\bm k})$ is independent of the
momentum $\bm{k}$ as in the conventional BCS  theory. 
The higher spin and multi-orbital natures of the system enable us to consider
various unconventional Cooper pairs even in the weakly correlated
case.

The gap function is expanded as
\begin{eqnarray}\label{eq:expand}
	\Delta = \sum_{i, \nu}\Delta_{i,\nu} \Phi_{i,\nu}\ , 
\quad
	\Phi_{i,\nu} = \frac{\varphi_i(\bm{J}) \otimes \sigma_\nu}{N},
\end{eqnarray}
where $\varphi_i(\bm{J})$ is a set of $4\times 4$ matrices spanning the
spin space,
$\sigma_{\nu}$ ($\nu=0,x,y,z$) is
the Pauli matrix in the orbital space, and 
$N\!=\!\sqrt{\mathrm{tr}[(\varphi_i \otimes \sigma_\nu)^2]/8}$ is the
normalization constant.
A convenient basis of $\varphi_i({\bm J})$ is the 
spherical harmonic $Y_{l,m}({\bm J})$, which is defined by $Y_{l,m\pm1}=\mp[J_{\pm},Y_{l,m}]$ with
$J_\pm=J_x\pm iJ_y$.  
In the present case, the spin of a Cooper pair is given by $3/2\otimes 3/2=0\oplus
1\oplus 2\oplus 3$, and hence, the azimuthal and magnetic
quantum numbers $l$ and $m$ in $Y_{l,m}$ assume the values $l=0,1,2,3$ and
$m=-l,\dots, l$.
Reconstructing the basis $Y_{l,m}$ for each $l$ in terms of
the representations of the $O$ 
group, we obtain $\varphi_i(\bm{J})$ in
Table \ref{table:Sclass}. 
Here, we use the $O$ group rather than $O_h$, as inversion acts on the spin space trivially as the identity operator.

\begin{table}[tb]
	\caption{Classification of spin basis for the $4\times4$ matrices according to $O$ point group symmetry. 
	The columns correspond to an
	irreducible representation of $O$ point group symmetry 
	with indices for the amplitude $l$ of coupled angular momentum for two $3/2$-spins, and the basis matrix.} \label{table:Sclass}
	\renewcommand{\arraystretch}{1.5}
	\begin{tabular}{l|c}
		\hline 
		$O^{(l)}$ & Basis $\varphi_i(\bm{J})$ \\
		\hline 
		\hline 
		$A_1^{(0)}$ & $1$ \\
		\hline 
		$T_1^{(1)}$ & $\{J_x, J_y, J_z\}$ \\
		\hline 
		$E^{(2)}$   & $\{2J_z^2\!-\!J_x^2\!-\!J_y^2, J_x^2\!-\!J_y^2\}$ \\
		$T_2^{(2)}$   & $\left\{ J_x J_y\!+\!J_y J_x,J_y J_z\!+\!J_z J_y, J_z J_x\!+\!J_x J_z \right\}$ \\
		\hline 
		$A_2^{(3)}$   & $J_xJ_yJ_z\!+\!J_yJ_zJ_x\!+\!J_zJ_xJ_y\!+\!J_zJ_yJ_x\!+\!J_xJ_zJ_y\!+\!J_yJ_xJ_z$ \\
		$T_1^{(3)}$   & $\{\tilde{J}_x,\ \tilde{J}_y,\ \tilde{J}_z \}$ \\
		$T_2^{(3)}$   & $\left\{J_yJ_xJ_y\!-\!J_zJ_xJ_z,\ J_zJ_yJ_z\!-\!J_xJ_yJ_x,\ J_xJ_zJ_x\!-\!J_yJ_zJ_y \right\}$ \\
		\hline
	\end{tabular}
	\renewcommand{\arraystretch}{1}
\end{table}

\begin{table}[b]
	\caption{Eight classes of the whole basis of the gap function according to $O_h$ point group symmetry. 
	The columns denote intraorbital or interorbital pairing, irreducible representation of $O_h$ point group symmetry, and the basis matrix
	given as the direct product of spin basis in $O$ representation and Pauli matrices for the orbital basis.} \label{table:Oh}
	\renewcommand{\arraystretch}{1.5}
	\begin{tabularx}{85mm}{c>{\centering} p{25mm}  p{15mm}c}
		\hline 
		Orbital & Parity & $O_h$ & Basis $\varphi_i(\bm{J})\otimes \sigma_\nu$ \\
		\hline 
			    & &$A_{1g}$ & $A_1^{(0)}\otimes \sigma_0$, $A_1^{(0)}\otimes\sigma_z$ \\
		Intra   & Even &$E_g$    & $E^{(2)}\otimes \sigma_0$, $E^{(2)}\otimes\sigma_z$ \\
				& &$T_{2g}$ & $T_2^{(2)}\otimes \sigma_0$, $T_2^{(2)}\otimes\sigma_z$ \\
		\hline 
				& &$A_{1u}$   & $A_1^{(0)}\otimes \sigma_x$ \\ 
				& &$A_{2u}$   & $A_2^{(3)}\otimes \sigma_y$ \\  
		Inter	& Odd &$E_u$   &  $E^{(2)}\otimes \sigma_x$ \\ 
				& &$T_{1u}$   &$T_1^{(1)}\otimes \sigma_y$, $T_1^{(3)}\otimes \sigma_y$ \\ 
				& &$T_{2u}$   & $T_2^{(2)}\otimes \sigma_x$, $T_2^{(3)}\otimes \sigma_y$ \\ 
		\hline
	\end{tabularx}
	\renewcommand{\arraystretch}{1}
\end{table}

Notably, a possible combination of $\varphi_i({\bm J})$ and $\sigma_\nu$ in 
the right-hand side of Eq.~(\ref{eq:expand}) is restricted by the
constraint 
\begin{eqnarray}\label{eq:parityconst}
	(\Delta({\bm k}) C_{2,\bm{y}})^T =-\Delta(-{\bm k}) C_{2,\bm{y}},
\end{eqnarray}
from the Fermi statistics of $\bm{c}_{\bm k}$.
As $\varphi_i({\bm J})$ for even (odd) $l$ satisfies 
$[\varphi_i(\bm{J})C_{2,\bm{y}}]^T= -\varphi_i(\bm{J})C_{2,\bm{y}}$
($[\varphi_i(\bm{J})C_{2,\bm{y}}]^T= +\varphi_i(\bm{J}) C_{2,\bm{y}}$),
it is combined only with the symmetric Pauli matrices   
$\sigma_0$, $\sigma_x$, and $\sigma_z$ (the anti-symmetric Pauli matrix
$\sigma_y$). 
Consequently, we have 28 different $\Phi_{i,\nu}$ in
Eq.~(\ref{eq:expand}).     
In the group theory, they are classified into eight irreducible
representations of the $O_h$ group, as shown in Table.~\ref{table:Oh}.

Here, we compare the possible gap functions in Table~\ref{table:Oh} with those in the superconductor in the single $J=3/2$ band.
In single $J=3/2$ systems such as half-Heuslers, 
only even-parity superconductivity corresponding to $A_{1g}$, $E_g$, and $T_{2g}$ in the Table~\ref{table:Oh} is  
possible within the ${\bm k}$ independent gap function~\cite{brydon2016},
because they do not have degrees of freedom of $\sigma_\nu$ associated with orbital parity.
In addition, we have odd-parity superconductivity in the Table~\ref{table:Oh}.
These odd-parity pairing states are expected to be topologically non-trivial~\cite{Sato2009, Sato2010}.
In the present system, these superconductivities 
are realized with a constant gap function
in contrast to the ${\bm k}$-dependent ones in single $J=3/2$ states~\cite{savary2017}.

\begin{figure}[t]
\includegraphics[width=85mm]{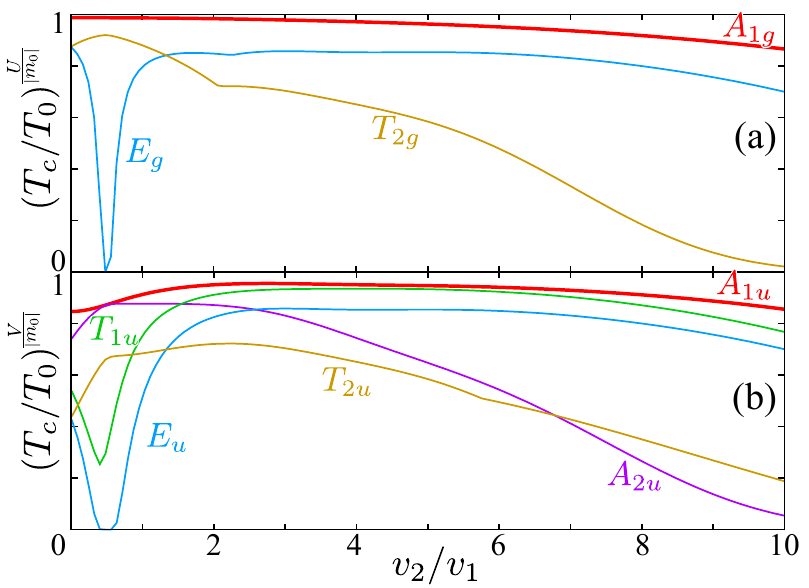}
\caption{Transition temperature for even- (a) and odd- (b) parity superconducting states. 
The unit temperature is $T_0=\frac{2e^\gamma \hbar \omega_0}{\pi k_{\mathrm{B}}}$ with Euler's constant $\gamma$ and cutoff frequency $\omega_0$.
Here, the parameters are $\mu=-2m_0$, $\alpha=6.25v_1^2/m_0$, and $\beta_i=0$.
}
\label{fig:Tc}
\end{figure}

\begin{figure}[b]
\includegraphics[width=85mm]{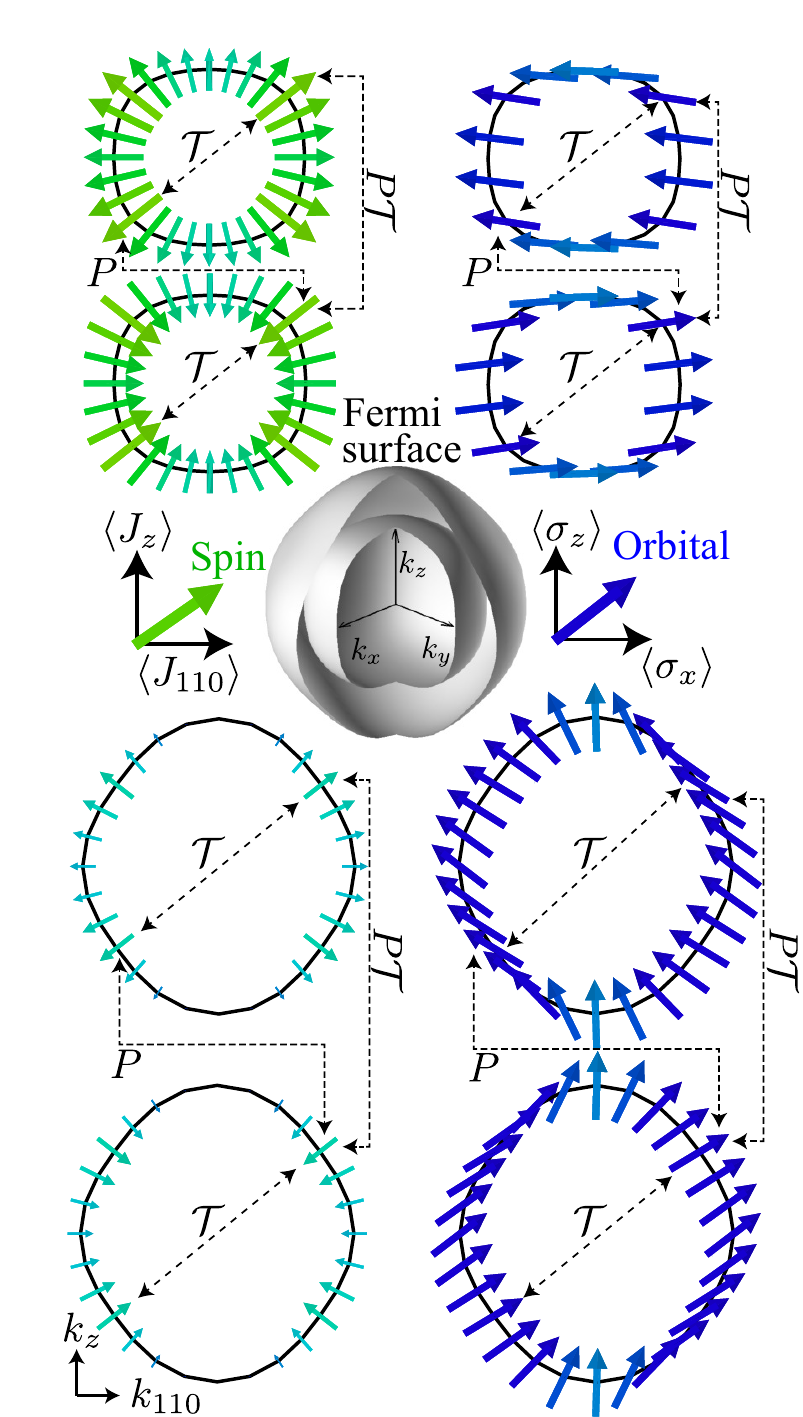}
\caption{(center) Fermi surfaces of antiperovskites. For $\mu<-m_0$, 
there are two Fermi surfaces.
The top and bottom panels show the spin and orbital textures of the inner and outer Fermi surfaces, respectively.
The green (blue) arrow represents the expectation value of the spin (orbit).
The dashed arrows indicate ${\cal T}$, $P$, and $P{\cal T}$ partners.
The parameters are $\mu=-2m_0$, $v_2=2.9v_1$, $\alpha=6.25v_1^2/m_0$, $\beta_i=0$, and $m_0>0$. 
}
\label{fig:texture}
\end{figure}

\subsection{Critical temperatures}\label{sec:Tc}
Now, we evaluate the superconducting transition temperature $T_{\mathrm{c}}$ for each representation in Table~\ref{table:Oh}.  
Here, we assume the following density-density pairing interactions, 
\begin{eqnarray}\label{eq:uv}
	H_{\mathrm{int}}\! =\! -\!\!\int\! d^3 r	\left[ U(n_{+}^2\!+\!n_-^2) + 2Vn_+n_- \right],
\end{eqnarray}
where $n_{\sigma_z}(\bm{r})=\sum_{j_z} \psi^\dag_{j_z,\sigma_z}(\bm{r})\psi_{j_z,\sigma_z}(\bm{r})$ is the density of the electron with orbital $\sigma_z$
defined by the field operator $\psi_{j_z,\sigma_z}(\bm{r})=\int d^3 k e^{-i\bm{k}\cdot\bm{r}}c_{j_z,\sigma_z,\bm{k}}$.

The transition temperature $T_{\mathrm{c}}$ is obtained by solving the linearized gap equation.
For each representation in Table~\ref{table:Oh}, 
we have~\cite{fu2010,hashimoto2016} (see also Appendix~\ref{sec:PTSC}),
\begin{eqnarray}\label{eq:tc}
	\mathrm{det}X(T_{\mathrm{c}}) = 0,
\end{eqnarray}
where the matrix element of $X(T_{\mathrm{c}})$ is 
\begin{eqnarray}\label{eq:Xapp}
	\!\!\!\!\!\!
	X_{\alpha\alpha'}(T_{\mathrm{c}}) &&\sim \delta_{\alpha\alpha'} +  
	\sum_{n} \frac{I(T_{\mathrm{c}})}{2}\!\! \int_{{k}_{\mathrm{F}}} d^2k' D_n(\bm{k}') {V}_\alpha \nn\\
	&&\times \!\!\!\sum_{n'=n,\bar{n}}\!\! \langle u_{n',\bm{k}'}|\Phi_{\alpha}|u_{n,\bm{k}'}\rangle\langle u_{n,\bm{k}'}|\Phi_{\alpha'} |u_{n',\bm{k}'}\rangle.
\end{eqnarray}
Here, $|{u}_{n,\bm{k}}\rangle$ is the normalized single-particle state for the Fermi surface of $n$-th band 
defined by $H_0(\bm{k})|{u}_{n \bm{k}}\rangle=0$. 
(Hereafter, all the ket vectors $|{u}_{n,\bm{k}}\rangle$ indicate the wave function of the normal state unless otherwise noted.)
The band index $\bar n$ represents the state 
$|{u}_{\bar n \bm{k}}\rangle = P\mathcal{T} |{u}_{n, \bm{k}}\rangle$, 
which is degenerated with $|{u}_{n \bm{k}}\rangle$.
$D_n(\bm{k}')=\frac{dk_F}{d\xi_n}$ is the density of states on the Fermi surface. 
The interaction is $V_{\alpha}=U$ ($V$) for intraorbital (interorbital) pairing. 
We also use $I(T_{\mathrm{c}})=\int_{-\hbar\omega_{0}}^{\hbar\omega_{0}} d\xi \frac{1}{2\xi}\tanh(\frac{\xi}{2k_{\mathrm{B}T_{\mathrm{c}}}})\sim \ln (\frac{2e^\gamma \hbar\omega_0}{\pi k_{\mathrm{B}}T_{\mathrm{c}}})$, Euler's constant $\gamma$,
cutoff frequency $\omega_0$, and Boltzmann constant $k_\mathrm{B}$. 
Here, $\alpha$ and $\alpha'$ specify a different basis $\varphi_i({\bm J})\otimes \sigma_{\nu}/N\equiv \Phi_{\alpha'}$ (if existing) in the representation.  
For instance, the $A_{1g}$ gap has two different bases, $A_1^{(0)}\otimes \sigma_0$ and $A_1^{(0)}\otimes \sigma_z$, and hence, $\alpha'$ runs $\alpha'=1,2$.

We {show} the numerical solutions of Eq.~(\ref{eq:tc}) {in Fig.~\ref{fig:Tc}}. 
The results for even-parity states and those for odd ones are shown
separately as 
$T_{\mathrm{c}}$'s of even- (odd-) parity states are independent of the
interorbital (intraorbital) pairing interaction $V$ ($U$).
It is observed that the spin-singlet $A_{1g}$ ($A_{1u}$) representation
exhibits the highest $T_{\mathrm{c}}$ among even- (odd-) parity states.

These results can be understood using the spin and orbital textures of
the Fermi surfaces in the normal state.
In Fig.~\ref{fig:texture}, we show the expectation values of $\langle
{\bm J}\rangle $ and $\langle {\bm \sigma}\rangle$ on the Fermi
surfaces with
respect to the single-particle state $|{u}_{n \bm{k}}\rangle$. 
The hole-doped system with $\mu<-m_0$
supports 
two Fermi surfaces around the $\Gamma$ point, and 
each Fermi surface has twofold degeneracy with different spin and
orbital textures, owing to the coexistence of inversion and time-reversal symmetries (see also Appendix~\ref{sec:PTSC}).  
The orbital texture $\langle {\bm \sigma}\rangle$ indicates mixing between orbitals with even and odd parity.
When $\langle {\bm \sigma}\rangle\parallel \hat{z}$, the
corresponding single-particle state satisfies $\sigma_z=\pm 1$, and hence, it
should be an orbital with either even or odd parity  and there is no orbital
mixing.
In contrast, when $\langle {\bm \sigma}\rangle \parallel \hat{x}$, 
orbitals with even and odd parity are fully mixed in equal weight.

\begin{figure}[b]
\includegraphics[width=85mm]{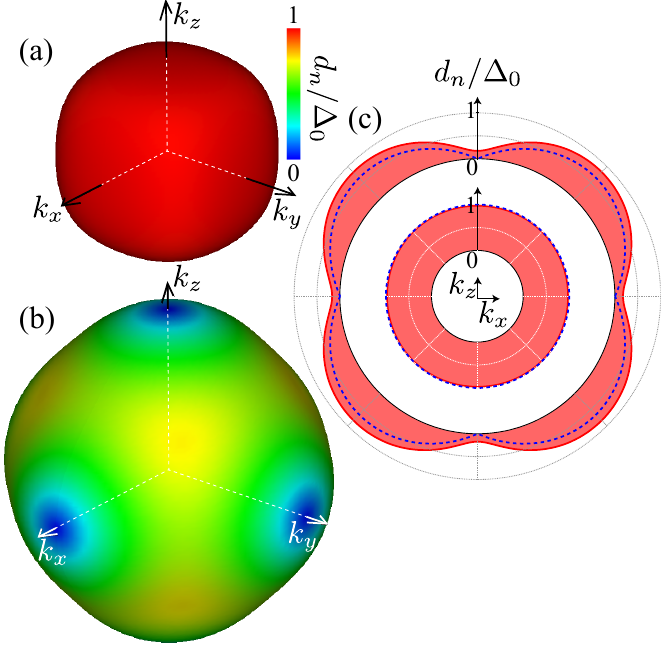}
\caption{Energy gap $d_{n}=\langle u_{n,{\bm k}}|\Delta |u_{n,{\bm k}}\rangle$ of $A_{1u}$ representation at the inner (a) and outer (b) Fermi surfaces for $v_{2}=3v_1$. 
Those for $v_{2}=3v_1$ (dashed) and $4v_1$ (solid line with filled lobe) on the $k_z k_x$ cut are displayed in (c).
}
\label{fig:gapstructure}
\end{figure}

First, consider the $A_{1g}$ state.
As it is a spin-singlet with the angular momentum $J\!=\!0$, the Cooper pair has an anti-parallel
spin configuration. Thus, it is formed by an electron and its
time-reversal partner, which have opposite spins.
In this case, they can form an intraorbital pairing without obstruction. 	
Indeed, as shown in Fig.~\ref{fig:texture}, as time-reversal does not
change the orbital, 
the electron and the
partner have the same $\langle {\bm \sigma} \rangle$. Consequently, 
the $A_{1g}$ state is naturally realized in the presence of
the intraorbital pairing interaction. 
It corresponds to the usual $s$-wave superconductivity, where the bulk is topologically trivial~[see also Sec.~\ref{sec:pairing}].

\begin{figure}[t]
\includegraphics[width=85mm]{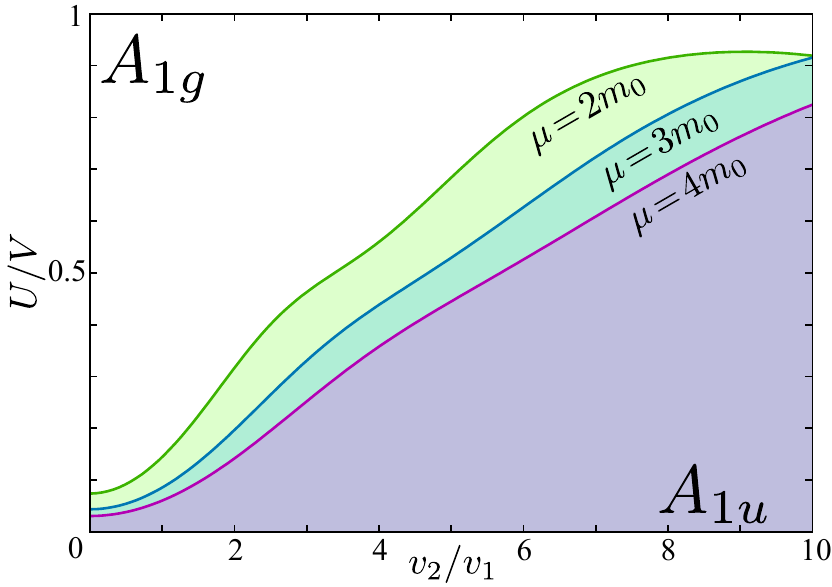}
\caption{Phase diagram of odd- and even-parity states spanned by $U/V$ and $v_2/v_1$.
The curves are the phase boundary for the typical values of chemical potential.
The other parameters are the same as in Fig.~\ref{fig:texture}.
}
\label{fig:Phdiag}
\end{figure}

Similarly, the $A_{1u}$ state is naturally realized in the
presence of the interorbital pairing interaction.
The $A_{1u}$ state is also spin-singlet with $J\!=\!0$, and hence, its Cooper pair is also
formed between an electron and its time-reversal partner. However,
for the interorbital pairing state, orbital mixing is
necessary, {which results in an interesting nodal structure of 
the gap depending on the model parameters.}
For instance, when $v_2= 3v_1$, the outer Fermi surface 
does
not show orbital mixing at the intersections with the primary
axis (i.e., the $k_{i=x,y,z}$ axis).
See Fig.~\ref{fig:texture}.
Correspondingly, there are gap nodes at the intersections in
Fig.~\ref{fig:gapstructure} (b).
{We will show later that this nodal structure is not accidental 
but has a topological origin.}
Note that, except for the special point 
$v_2= 3v_1$, $v_1/2$, and $-v_1/3$, the $A_{1u}$ state is full-gap as shown in Fig.~\ref{fig:gapstructure} (c).
In Fig.~\ref{fig:Phdiag}, we show the phase diagram obtained through our calculation.
It is observed that {the} $A_{1u}$ {phase} is realized when $V$ is much larger than $U$.

\section{Topological superconductivity in the $A_{1u}$ state}\label{sec:A1u}
In this section, 
we study the topological properties of the $A_{1u}$ {($J=0$)} superconducting state,
in terms of the topological indices and numerically obtained surface states.

\subsection{3D winding number}~\label{sec:W3D}
 
We first examine the topological properties independent of the crystal symmetry.
For the time-reversal symmetric superconductivity generally described by the BdG Hamiltonian~(\ref{eq:bdg}),
we have chiral symmetry, 
\begin{eqnarray}\label{eq:chs}
 \Gamma H(\bm{k}) \Gamma=-H(\bm{k}) \hbox{ with } \Gamma=\tau_y,
\end{eqnarray}
which corresponds to the combination of time-reversal and particle-hole symmetries [see Appendix~\ref{sec:PTSC} for more details].
As these symmetries do not depend on any particular crystal structure, 
the chiral symmetry specifies the most general symmetry-protected topological number~\cite{schnyder2008}.
In the present case, the general topological number is the 3D winding number,
{\begin{eqnarray}\label{eq:w3d}
	w_{\mathrm{3D}} = \int \frac{d^3k}{48\pi^2} 
\epsilon^{\alpha\beta\gamma}
\mathrm{tr}
\left[\Gamma
H^{-1} \partial_\alpha H 
H^{-1} \partial_\beta H 
H^{-1} \partial_\gamma H 
\right].
\nonumber\\
\end{eqnarray}}%
Here, we calculate $w_{\mathrm{3D}}$ for the $A_{1u}$ state with $\Delta=\Delta_0 \sigma_x$. 
In particular, we consider the weak pairing limit, $\Delta_0/\mu\ll 1$, similar to ordinary superconductors. 

First, we numerically evaluate $w_{\rm 3D}$.
It is observed that, when the absolute value of the chemical potential $|\mu|$ is less than the critical value $\mu_c$,  
\begin{eqnarray}
\mu_c=\sqrt{m_0^2+\Delta_0^2} 
\end{eqnarray}
$w_{\rm 3D}$ is zero i.e.,
\begin{eqnarray}\label{eq:W3D}
	w_{\mathrm{3D}}(|\mu|<\mu_c)= 0.
\end{eqnarray}
At $|\mu|=\mu_c$, the BdG Hamiltonian in Eq.~(\ref{eq:bdg}) becomes gapless at $k=0$; thus 
topological phase transition takes place at $\mu=\mu_c$, then 
for $|\mu|>\mu_c$, we have
\begin{eqnarray}\label{eq:W3D}
	w_{\mathrm{3D}}(|\mu|>\mu_c)= 
	\left\{\begin{array}{rl}
		-2\,\mathrm{sign}(v_1) &\mbox{for $\frac{v_2}{v_1} < -\frac{1}{3}$} \\
		4\,\mathrm{sign}(v_1) & \mbox{for $-\frac{1}{3}< \frac{v_2}{v_1}<\frac{1}{2}$}\\
		-4\,\mathrm{sign}(v_1) & \mbox{for $\frac{1}{2}<\frac{v_2}{v_1}<3$} \\
		 2\,\mathrm{sign}(v_1) &\mbox{for $\frac{v_2}{v_1} > 3$}
	\end{array}\right. .\nonumber\\
\end{eqnarray}
For $|\mu|>\mu_c$, the system supports two Fermi surfaces (inner and outer Fermi surfaces) around the $\Gamma$ point, as illustrated in Fig.~\ref{fig:texture}.  
When $w_{\rm 3D}$ changes at $v_2/v_1=-1/3, 3$ $(v_2/v_1=1/2)$, six (eight) point nodes appear on the outer Fermi surface, 
each of which contributes to the change of $w_{\rm 3D}$ by 1 ($-1$) with the increase in $v_2$. 
These results indicate that the $A_{1u}$ state hosts nontrivial topological superconductivity when $|\mu|>\mu_c$. 
In particular, it supports a characteristic higher winding number $|w_{\rm 3D}|\ge 1$.
As discussed immediately below, the higher winding number of the topological superconductivity is a direct consequence of the higher spin of the present system. 

Here, we consider the origin of the higher winding number.
Usually, the pairing interaction is too weak to mix the energetically separated bands. 
In this weak pairing case, the topological number of the superconductor 
is attributed to that of each band forming Fermi surfaces,
where the electron and hole states are degenerated.
The contribution from each band remains unchanged unless 
the gap of the system closes or the Fermi surfaces contact each 
other.\footnote{In the strong pairing case, the large gap function may mix the contributions from each band, 
but their summation is conserved.}
In the present case, we have~\cite{qi2010}
\begin{eqnarray}
w_{\rm 3D}=\frac{1}{2}\sum_n{\rm sign}\left[v^n_{\rm F}\right]{\rm sign}\left[d_n\right] \nu_{\rm Ch}^n,
\label{eq:FSformula}
\end{eqnarray}
where $n$ is summed for all band indices forming disconnected Fermi surfaces, 
$v_{\rm F}^n$ is the Fermi velocity on the $n$-th Fermi surface, and
$d_n$ is the expectation value of the gap function 
\begin{eqnarray}
d_n=\langle u_{n,{\bm k}}|\Delta|u_{n,{\bm k}}\rangle, 
\end{eqnarray}
the ket vector $|u_{n,{\bm k}}\rangle$ is the solution of Bloch equation $H_0(\bm{k})|u_{n,\bm{k}}\rangle=0$ on the $n$-th Fermi surface,
and $\nu_{\rm Ch}$ is the first Chern number of the Fermi surface,
\begin{eqnarray}
\nu^n_{\rm Ch}= -i\int_{k=k_{\mathrm{F}}^{n}} d^2k' \epsilon^{\alpha\beta} \langle \partial_{k_{\alpha}'} u_{n,\bm{k}}| \partial_{k_\beta'} u_{n,\bm{k}} \rangle 
\end{eqnarray}
where $k_\alpha'$ represents the two-dimensional momenta on the Fermi surface. 
Here, we consider Kramers partners separately in the summation of $n$.
See Appendix~\ref{sec:PTSC} for more details.

The relation between the $J=3/2$ spin and the higher winding number becomes evident 
in the spherical symmetric case at $v_2=0$.
In this case, the solution of the Bloch equation $H_0(\bm{k})|u_{j_z,\pm,\bm{k}}\rangle = \xi_{j_z,\pm}|u_{j_z,\pm,\bm{k}}\rangle$ for the normal state is
given by
\begin{eqnarray}\label{eq:wfnsph}
	|u_{j_z,\pm,\bm{k}}\rangle = R(\hat{\bm k},\bm{J})\left|j_z\right> \otimes \left|\Psi_{\pm}(\rho_k)\right>_\sigma.
\end{eqnarray}
Here, $R(\hat{\bm k}, \bm{j})=e^{-iJ_z\phi_k}e^{-iJ_y\theta_k}$ 
with the polar and azimuthal angles ($\theta_k$, $\phi_k$) of $\bm{k}$ is the rotation matrix that diagonalizes the spin-dependent part of $H_0({\bm k})$ as 
$R^\dag(\hat{\bm k},\bm{J}){\bm k}\cdot{\bm J}R(\hat{\bm k},\bm{J})=k J_z$ , 
and $\left|j_z\right>\in\{\left|\pm 3/2\right>, \left|\pm 1/2\right>\}$ is the eigenstate of $J_z$. 
The orbital-dependent part of the wave functions is given by
\begin{eqnarray}~\label{eq:uOrb}
	\left|\Psi_{\pm}(\rho_k)\right>_\sigma = e^{-i\sigma_y \rho_k}{\left|\pm\right>},
\end{eqnarray}
with $\rho_k=\frac{1}{2}\arctan[v_1kj_z /m(k)]$ and the eigenstate $\left|\pm\right>$ of $\sigma_z$.
It is observed that the states $|u_{j_z,+,\bm{k}}\rangle$ ($|u_{j_z,-,\bm{k}}\rangle$) 
yield the electron (hole) branches of the normal spectra, and hence, they define the Fermi surfaces 
for $\mu> \mu_c$ ($\mu< -\mu_c$).
Therefore, Eq. (\ref{eq:FSformula}) leads to   
\begin{eqnarray}
w_{\rm 3D}=\sum_{j_z} w_{\rm 3D}^{j_z},
\end{eqnarray}
with
\begin{eqnarray}\label{eq:W3Dsph}
	w_{\mathrm{3D}}^{j_z} = 
\pm \frac{1}{2}{\rm sign}(d_{j_z,\pm} ) 
\nu^{j_z}_{\mathrm{Ch}}, 
\end{eqnarray}
where the double sign $\pm$ corresponds to the case with $\mu\gtrless\pm\mu_c$, $d_{j_z,\pm} = \langle u_{j_z,\pm,\bm{k}}|\Delta_0\sigma_x |u_{j_z,\pm,\bm{k}}\rangle$, 
and
$\nu^{j_z}_{\mathrm{Ch}}$ is the first Chern number of $|u^\pm_{j_z, {\bm k}}\rangle$ on each Fermi surface. 
Equation~(\ref{eq:wfnsph}) yields 
\begin{eqnarray}
{\rm sign}(d_{j_z,\pm,\bm{k}}) = \pm {\rm sign}(j_z v_1).
\end{eqnarray}
Using the polar coordinates of the Fermi surface, we can also evaluate $\nu_{\rm Ch}^{j_z}$ analytically,
\begin{eqnarray}
	\nu^{j_z}_{\mathrm{Ch}} &=& \frac{1}{2\pi} \int d\theta_k d\phi_k 
	\left(-i \left\langle\frac{\partial u^\pm_{j_z,\bm{k}}}{\partial \theta_k}\Bigg|\frac{\partial u^\pm_{j_z,\bm{k}}}{\partial \phi_k}\right\rangle+ \mathrm{h.c.}\right)
	\nonumber\\
	&=&2 j_z.
\end{eqnarray}
Thus, we have
\begin{eqnarray}
w_{\rm 3D}^{j_z}={\rm sign}(v_1)|j_z|,
\end{eqnarray}
both for $\mu>\mu_c$ and $\mu<\mu_c$.

The last expression indicates that a higher spin provides a higher winding number.
Using this expression, we can also evaluate the winding number for each Fermi surface.
For $v_2=0$, the outer (inner) Fermi surface consists of $j_z=\pm 1/2$ ($j_z=\pm 3/2$) components, and hence, we obtain
\begin{eqnarray}
&& w_{\mathrm{out}} = w_{\mathrm{3D}}^{1/2}+w_{\mathrm{3D}}^{-1/2}= {\rm sign}(v_1), 
\nonumber\\
&& w_{\mathrm{in}} = w_{\mathrm{3D}}^{3/2}+w_{\mathrm{3D}}^{-3/2}= 3 {\rm sign}(v_1),
\label{eq:woutwin}
\end{eqnarray}
where $w_{\rm out}$ ($w_{\rm in}$) denotes the winding number of the outer (inner) Fermi surface.
These winding numbers retain the same values for $-1/3<v_2/v_1<1/2$ as either a gap-closing or a contact of Fermi surfaces does not occur until $v_2/v_1$ reaches the boundary of the region at $v_2/v_1=1/2$ or $-1/3$.
Here, we note that the total winding number $w_{\rm out}+w_{\rm in}$ reproduces the numerical result of Eq.~(\ref{eq:W3D}) in the same region of $-1/3<v_2/v_1<1/2$.

Using the above result, we can also identify the winding numbers $(w_{\rm out}, w_{\rm in})$ for the other regions of $v_2/v_1$.
First, to evaluate them in the region $1/2<v_2/v_1<3$, we use a {``}duality'' relation. 
As explained in Appendix~\ref{sec:duality}, 
a unitary transformation maps the parameter $(v_1, v_2)$ to $(-3v_1/5-4v_2/5, -4v_1/5+3v_2/5)$.
This transformation exchanges the region of $-1/3<v_2/v_1<1/2$ with that of $1/2<v_2/v_1<3$, retaining the energy spectra and reversing the winding numbers. 
Thus, combining the duality relation with Eq.~(\ref{eq:woutwin}), we obtain
 \begin{eqnarray}
w_{\rm out}=-{\rm sign}(v_1),
\quad
w_{\rm in}=-3{\rm sign}(v_1),
\end{eqnarray}
for $1/2<v_2/v_1<3$.
Subsequently, to determine the winding numbers for the remaining two regions $v_2/v_1<-1/3$ and $v_2/v_1>3$, we use the {properties} of the topological phase transitions at $v_2/v_1=-1/3$ and $v_2/v_1=3$.  
As mentioned above, 
at the topological phase transitions, six point nodes appear on the outer Fermi surface, 
which change $|w_{\rm out}|$ by 6~\footnote{We observe that $|w_{\rm out}|$ changes only by 2 at $v_2/v_1=1/2$ although there appear eight point nodes on the outer Fermi surface. This is because the inner and outer Fermi surfaces contact each other at $v_2/v_1=1/2$, and hence, they exchange their winding numbers.}.
Provided  also that the total winding number $w_{\rm out}+w_{\rm in}$ reproduces Eq.~(\ref{eq:W3D}), we can uniquely determine the winding numbers as
\begin{eqnarray}
w_{\rm out}=-5{\rm sign}(v_1),
\quad
w_{\rm in}=3{\rm sign}(v_1),
\end{eqnarray} 
for $v_2/v_1<-1/3$, and 
\begin{eqnarray}
w_{\rm out}=5{\rm sign}(v_1),
\quad
w_{\rm in}=-3{\rm sign}(v_1),
\end{eqnarray} 
for $v_2/v_1>3$. 
In both cases, the Fermi surfaces have {characteristic} higher winding numbers.

\subsection{Mirror Chern numbers}\label{sec:mCh}
The present system has the $O_h$ group crystalline symmetry, which provides more
detailed topological information~\cite{chiu2013,
morimoto2013,shiozaki2014, shiozaki2016, shiozaki2017}.
Specifically, we first consider
the horizontal (or equivalently vertical) mirror reflection $M_x$ with
respect to the $k_x$-axis.
\begin{figure}[bt]
\includegraphics[width=85mm]{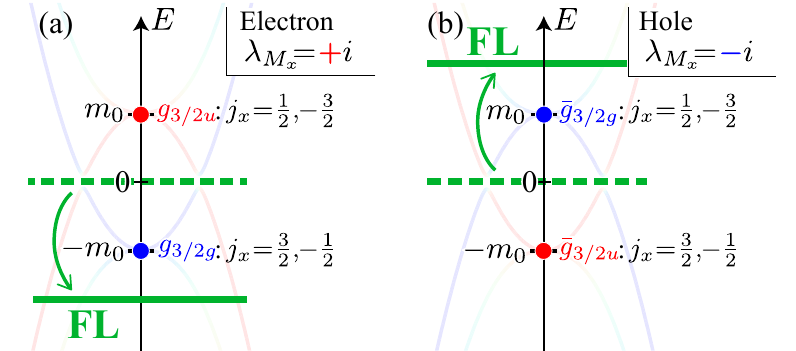}
\caption{ Level structures and their angular momentum in the 
$\left<\mathcal{M}_x\tau_z\right>=i$ sector at the $\Gamma$ point. (a) and (b) indicate electron and hole sectors, respectively.
}
\label{fig:level}
\end{figure}

The $A_{1u}$ gap function is odd under the mirror reflection.
In this case, the mirror reflection operator for the BdG Hamiltonian is
given by $M\tau_z$, where $M=PC_{2,\bm{n}}$ is the mirror operator for the normal
Hamiltonian $H_0({\bm k})$, and ${\bm n}$ is the normal vector of the reflection plane~\cite{ueno2013}.
On the mirror-invariant plane $\bm{k}\cdot\bm{n}=0$, the BdG Hamiltonian 
can be block diagonal in the diagonal basis of $M\tau_z$. Thus, the mirror
Chern number $\tilde{\nu}_{\sigma_{h,d}}$ for the BdG Hamiltonian is defined 
in the same manner as $\nu_{\sigma_{h,d}}$ in Sec.\ref{sec:normal}.

First, let us evaluate $\tilde{\nu}_{\sigma_h}$ in the limit $\mu=0$ and $\Delta_0 = 0$ 
for the mirror reflection with respect to three equivalent horizontal planes: $x=0$, $y=0$, or $z=0$.  
In this limit, the BdG Hamiltonian reduces to $H_0({\bm k})\tau_z$, and hence,
$\tilde{\nu}_{\sigma_h}$ can be evaluated as
$\tilde{\nu}_{\sigma_h}=2\nu_{\sigma_h}$.  
Therefore, from Eq.~(\ref{eq:chh0+}), we have
\begin{eqnarray}\label{eq:Chh_smallmu}
	\tilde{\nu}_{\sigma_h}(\mu\!=\!0)= 4 \mathrm{sign}(b_{3/2}b_{1/2}).
\end{eqnarray}
For small $\Delta_0$ and $\mu$, $\tilde{\nu}_{\sigma_h}$ retains the same
value unless the gap of the system closes.

When $|\mu|=\mu_c$, the gap of the system closes at the $\Gamma$ point, and
$\tilde{\nu}_{\sigma_h}$ changes.
Using the fourfold rotation symmetry around the $k_x$ axis, 
we obtain the following useful formula to calculate
$\tilde{\nu}_{\sigma_h}$~\cite{kobayashi2015}.
\begin{eqnarray}
	\tilde{\nu}_{\sigma_h}= \sum_{j_x}N_{j_x,+}(\Gamma) j_x \mod 4,
\end{eqnarray}
where $N_{j_x,+}(\Gamma)$ is the number of negative energy states with
the spin $j_x$ at the $\Gamma$ point in the eigensector of $M_x\tau_z$ 
with the eigenvalue $\lambda_{{M}_x\tau_z}=\pm i$.
In the weak coupling limit $\Delta_0\rightarrow 0$, 
the particle and hole sectors are decoupled, so that $\lambda_{{M}_x\tau_z}=\lambda_{{M}_x}\lambda_{\tau_z}$. 
In this case, a negative energy
state in the $\lambda_{M_x\tau_z}=i$ sector reduces to either an electron state ($\lambda_{\tau_z}=+1$) in the
$\lambda_{M_x}=i$ sector or a hole state ($\lambda_{\tau_z}=-1$) in the $\lambda_{M_x}=-i$
sector below the Fermi level.
Therefore, $N_{j_x,+}(\Gamma)$ can be evaluated as the number of
these electron and hole states. 
Thus, as summarized in Fig.~\ref{fig:level}, 
when $|\mu|$ exceeds $m_0$, the $j_x=3/2, -1/2$ electron ($j_x=1/2,
-3/2$ hole) bands in the $\lambda_{M_x}=i$ ($\lambda_{M_x}=-i$) sector move above (below) the Fermi level at $\Gamma$. 
Therefore, the mirror Chern number jumps by 
\begin{eqnarray}\label{eq:delnu}
	\Delta \tilde{\nu}_{\sigma_h} &=& -\!\left(\frac{3}{2}\!-\!\frac{1}{2}\right) + \left(\frac{1}{2}\!-\!\frac{3}{2}\right) \mod 4 \nn\\
	&=& -2 \mod 4
\end{eqnarray}
We can also evaluate $\tilde{\nu}_{\sigma_h}$ numerically as
\begin{eqnarray}\label{eq:Chh_largemu}
	\tilde{\nu}_{\sigma_h}(|\mu|>\mu_c)
= 2 \mathrm{sign}(b_{3/2}b_{1/2}),
\end{eqnarray}
which is consistent with Eq.~(\ref{eq:delnu}).

\begin{figure}[b]
\includegraphics[width=85mm]{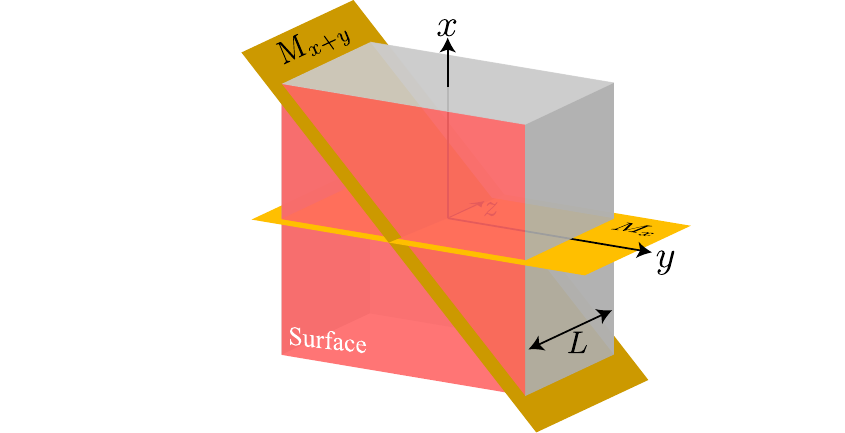}
\caption{System geometry of the slab system.
}
\label{fig:slab}
\end{figure}

According to the bulk-edge correspondence, the non-zero mirror Chern
number indicates the existence of surface states. 
Here, we calculate the energy spectrum in the slab geometry
in Fig.~\ref{fig:slab} with a finite size $L$ along the $z$ axis. 
This system has horizontal (diagonal) mirror reflection symmetry with respect to
the $x$ ($x+y$) direction even in the presence of the surface.
Along the high symmetric line $k_x=0$ ($k_x+k_y=0$), 
we numerically solve the BdG equation for each mirror sector by replacing 
the momentum $k_z\rightarrow -i\partial_z$, and using the Gauss--Lobatto expansion method~\cite{mizushima2010}.

\begin{figure}[t]
\includegraphics[width=85mm]{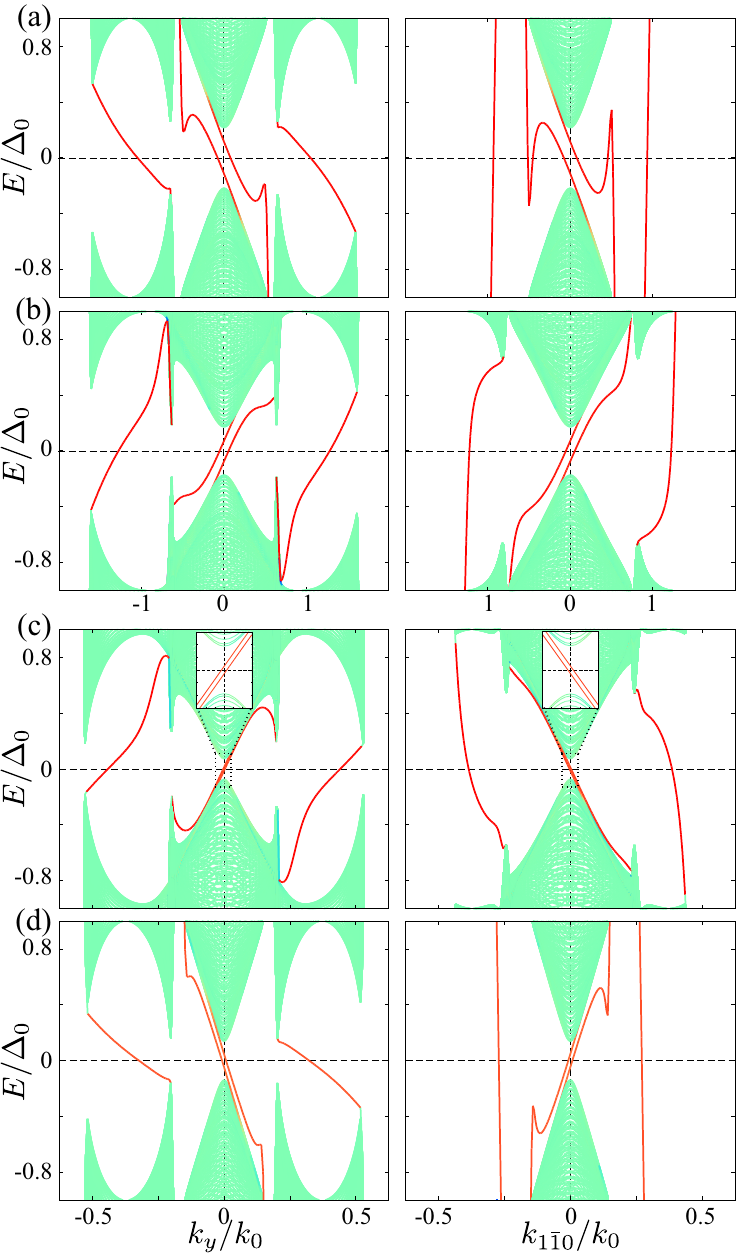}
\caption{Energy spectrum of the slab in different topological phases of a small $|\mu|$ regime along $k_x$ (left) and $k_{1\bar{1}0}=(k_x-k_y)/\sqrt{2}$ (right) directions.  
For better illustration, we display the surface states localized only at $z=-L/2$ highlighted by the red curves. 
The momentum unit is $k_0=m_0/v_1$. Insets of (b) are magnified images of the center of the Brillouin zone and close to $E=0$.
Here, $\Delta_0=0.025 m_0$, $\mu=-0.75 m_0$, $v_2/v_1=-0.5$ (a), $-0.2$ (b), $2.5$ (c), and $4$ (d), 
$\alpha= 0.64v_1^2/m_0$ (a,b) and $6.25v_1^2/m_0$ (c,d).
}
\label{fig:mu0.75}
\end{figure}

\begin{figure}[t]
\includegraphics[width=85mm]{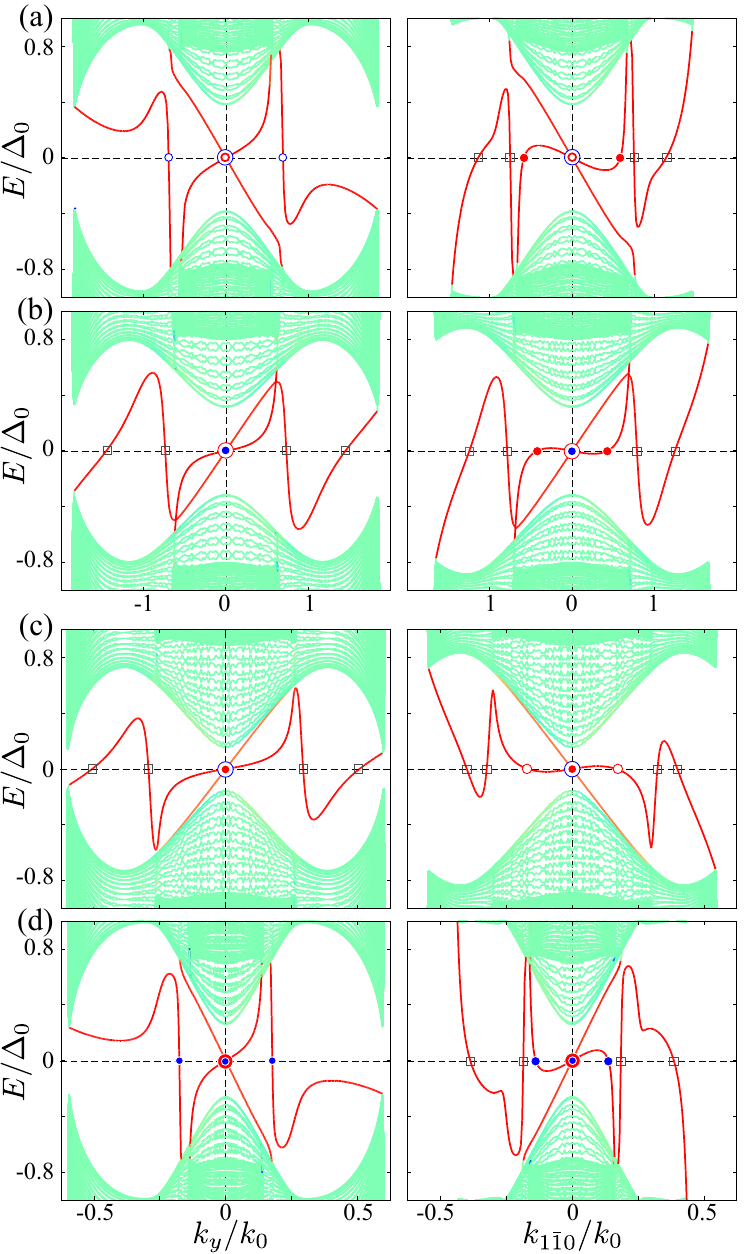}
\caption{
Energy spectrum of the slab in different topological phases for a large $|\mu|$ regime, $\mu=-1.25 m_0$. 
The red and blue symbols denote the zero-energy state of helical Majorana fermions 
originating from the outer and inner Fermi surfaces, respectively. 
The closed and open ones indicate the positive and negative helicities, respectively . 
{The open square symbols indicate pairs of helical Majorana fermions with opposite helicities, 
which can annihilate via continuous deformation.}
The other conditions are the same as in Fig.~\ref{fig:mu0.75}.
}
\label{fig:mu1.25}
\end{figure}

As shown in left column of Fig.~\ref{fig:mu0.75}, when $|\mu|<\mu_c$, the
spectrum along the $k_x=0$ line exhibits four branches of surface states. 
The number of branches coincides with $|\tilde{\nu}_{\sigma_h}|$.
Furthermore, the surface states change the chirality (or the
connectivity with the bulk states) when
$\tilde{\nu}_{\sigma_h}$ in the bulk changes the sign. 
For $-1/3<v_1/v_2<3$ [see Fig.~\ref{fig:mu0.75}(b) and (c)], each surface state connects the bulk bands upward
with the increase in $k_y$,
but for $v_1/v_2<-1/3$ or for $v_1/v_2>3$ [see Fig.~\ref{fig:mu0.75}(a) or (d)],
it connects them downward.
A similar bulk-edge correspondence holds for the higher-doped phase $|\mu|>\mu_c$ as shown in Fig.~\ref{fig:mu1.25}.
In this case, when the surface state cuts the zero energy upward (downward) with the increase in $k_y$,
it accumulates the mirror Chern number by 1 $(-1)$.
The total mirror Chern number counted from Fig.~\ref{fig:mu1.25}
is consistent with Eq.~(\ref{eq:Chh_largemu}).

Similarly, the diagonal mirror Chern number $\tilde{\nu}_{\sigma_d}$
with respect to the $k_x+k_y$ direction can be defined 
for the BdG {Hamiltonian}.
For $\mu\!=\!0$, it is evaluated as twice the value of that in
Fig.~\ref{fig:Ndirac}(f), which persists as long as $\mu$ reaches
the critical value $\mu_c$,
\begin{eqnarray}\label{eq:Chd_smallmu}
	\tilde{\nu}_{\sigma_d}(|\mu|<\mu_c)=
	\left\{\begin{array}{l}
	+4 \hbox{ for } -\frac{1}{3}<\frac{v_2}{v_1}<\frac{1}{2}, \\
	-4 \hbox{ for } \frac{1}{2}<\frac{v_2}{v_1}<3, \\
	0 \hbox{ otherwise }. \\
	\end{array}\right. 
\end{eqnarray}
In addition, for $|\mu|>\mu_c$, we numerically determine that 
\begin{eqnarray}\label{eq:Chd_largemu}
	\tilde{\nu}_{\sigma_d}(|\mu|\!>\!\mu_c)=
	\left\{\begin{array}{l}
	+2 \hbox{ for } -\frac{1}{3}<\frac{v_2}{v_1}<\frac{1}{2}, \\
	-2 \hbox{ for } \frac{1}{2}<\frac{v_2}{v_1}<3, \\
	0 \hbox{ otherwise }. \\
	\end{array}\right. 
\end{eqnarray}
The quasi-particle spectra along the mirror-invariant line $k_x+k_y=0$ for the slab geometry
are shown in right columns of Figs.~\ref{fig:mu0.75} and \ref{fig:mu1.25}, 
all of which are consistent with the above values of the diagonal mirror Chern number.
In particular, for $\mu=-0.75 m_0$, 
the surface dispersions of Figs.~\ref{fig:mu0.75}(a) and (d) cut 
the zero energy upward as many times as downward, from right to left, 
and hence, there is no topological protection.

In Fig~\ref{fig:mu1.25}, all the surface states for
$|\mu|> \mu_c$ pass through the high symmetry point $k_y=0$
($k_{1\bar{1}0}=0$) at $E=0$.
The origin of this property will be explained in the next section.

\begin{figure*}[tb]
\includegraphics[width=170mm]{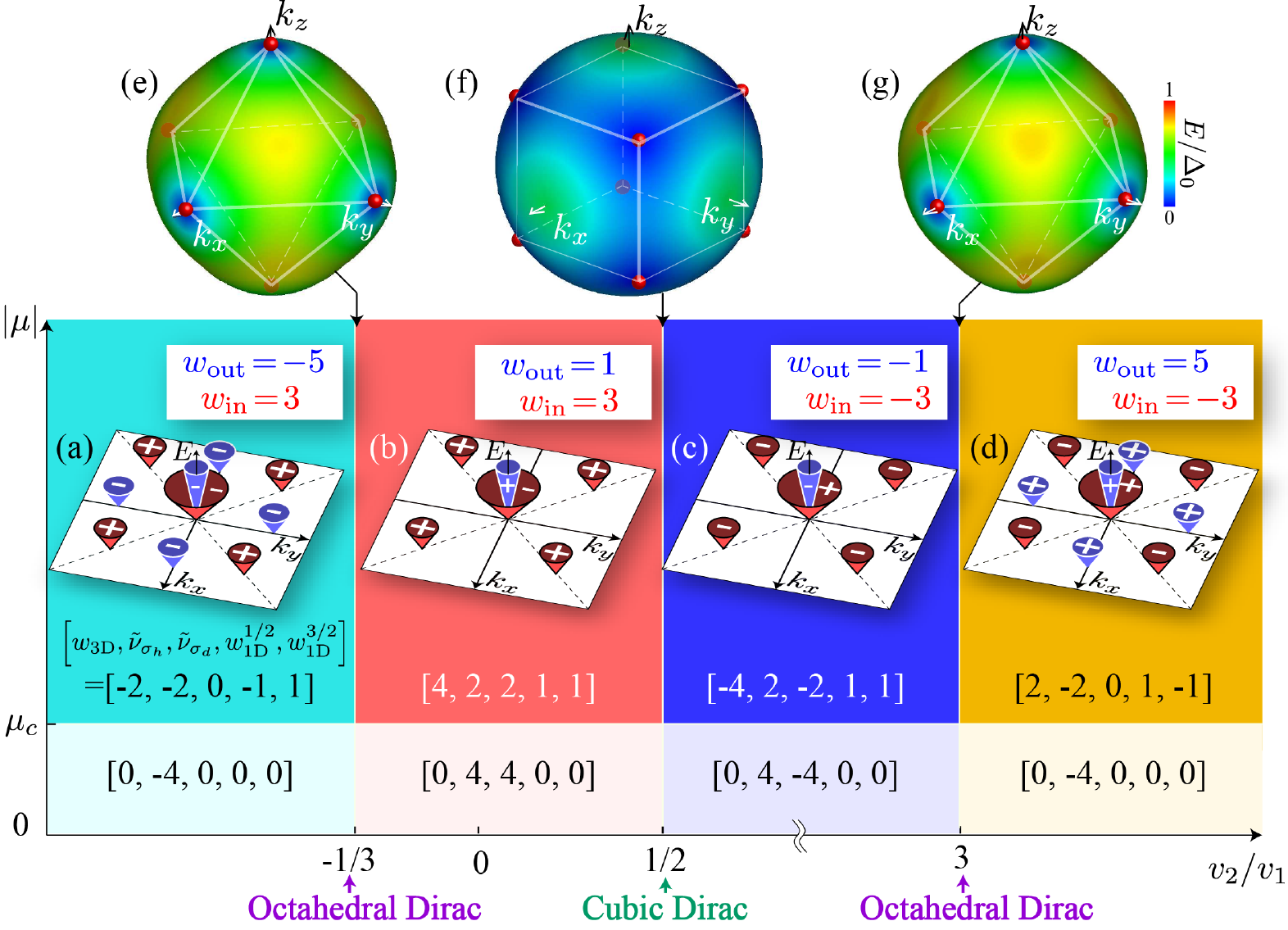}
\caption{Topological phase diagram of the $A_{1u}$ pairing state for $v_1>0$. 
The $A_{1u}$ state may host the 3D winding number $w_{\mathrm{3D}}$ as a time-reversal invariant superconductor. From the $O_h$ group symmetry, it also supports the two different mirror Chern numbers, $\tilde{\nu}_{\sigma_h}$ and $\tilde{\nu}_{\sigma_d}$, and the rotation 1D winding number $w_{\mathrm{1D}}^{j_z}$. 
If we consider $v_1<0$, the winding numbers $w_{\mathrm{3D}}$ and $w_{\mathrm{1D}}^{j_z}$ change their signs .
The insets indicate the location of a helical Majorana fermion in the surface Brillouin zone, 
which carries by the winding number  $\pm 1$.
The cones with blue and red in the $|\mu|>\mu_c$ phase show the contributions from the outer and inner Fermi surfaces, respectively. 
These colors correspond to the symbols in Fig.~\ref{fig:mu1.25}.
The gap structures on the outer Fermi surface 
(e) $v_2=3v_1$, (f) $v_2=v_1/2$, and (g) $v_2=v_1/2$.
The red dots indicate the point node of the superconducting gap. 
Here, $\mu=-2 m_0$, $\beta_i=0$, $m_0>0$, and $\alpha=6.25 v_1^2/m_0$ for (e,f) and $\alpha=0.64 v_1^2/m_0$ for (g).
}
\label{fig:TPdiag}
\end{figure*}

\subsection{1D winding numbers}\label{sec:W1D}
The system also has fourfold rotation symmetry, from which we can
define another topological invariant. 
The $A_{1u}$ gap function is invariant under the fourfold
rotation along the primary axis, and hence, the BdG Hamiltonian trivially realizes
the fourfold symmetry as 
\begin{eqnarray}
\tilde{C}_{4,{\bm z}}
H(k_y, -k_x, k_z)\tilde{C}_{4,{\bm z}}^{-1}=H({\bm k}),
\quad  \tilde{C}_{4,{\bm z}}=C_{4,{\bm z}}\tau_0.
\end{eqnarray}
On the primary axis $k_x=k_y=0$, the BdG Hamiltonian is block
diagonal in the eigenbasis of $\tilde{C}_{4,{\bm z}}$ with the eigenvalue $\lambda_{\tilde{C}_{4,\bm{z}}}=e^{-i\frac{j_z}{2}\pi}$,
\begin{eqnarray}\label{eq:bdgmz}
	H_{j_z}(k_z) = H_{0,j_z}(k_z)\tau_z + \Delta_0\sigma_x \tau_x,  
\end{eqnarray}
where $H_{0,j_z}(k_z)$ is given in Eq.~(\ref{eq:mzsector}). We also observe that each sector $H_{j_z}(k_z)$ retains chiral symmetry $\Gamma=i{\cal T}{\cal C}$ in Eq.~(\ref{eq:chs}), as it holds that
$[\Gamma, \tilde{C}_{4,\bm{z}}]=0$. 
Thus, on the basis where $\Gamma$ is diagonal, 
$H_{j_z}(k_z)$ takes the form 
\begin{eqnarray}
	H_{j_z} = 
	\left(\begin{array}{cc}
	0 & h_{j_z}(k_z) \\
	h_{j_z}^\dag(k_z) & 0
	\end{array}\right).
\end{eqnarray}
Thus, we can introduce the 1D winding number 
\begin{eqnarray}
	w^{j_z}_{\rm 1D} = \frac{1}{2\pi} \mathrm{Im} \left\{\int dk_z  \partial_{k_z} \ln[\det (h_{j_z})]\right\},
\end{eqnarray}
with
\begin{eqnarray}
	\det (h_{j_z})\!=\! \mu^2\!-\!m_{j_z}^2(k_z)\!+\!(b_{j_z} k_z)^2\!-\!\Delta^2_0\!+\!2i b_{j_z}k_z\Delta_0,
\end{eqnarray}
where $b_{j_z}$ is given by Eqs.~(\ref{eq:b32}) and (\ref{eq:b12}).
It can be calculated analytically as 
\begin{eqnarray}
	w_{\mathrm{1D}}^{j_z}=
	\left\{\begin{array}{ll}
		0 & \hbox{for } |\mu| < \mu_c \\
		\mathrm{sign}(b_{j_z}) &\hbox{for } |\mu| > \mu_c
	\end{array}\right. .
\end{eqnarray}
It is also observed that the total winding number vanishes, $\sum_{j_z}
w_{\mathrm{1D}}^{j_z}=0$, because of $b_{-j_z}=- b_{j_z}$, and 
the 1D winding number is related to the mirror {C}hern
number as 
\begin{eqnarray}
	2w_{\rm 1D}^{3/2}w_{\rm 1D}^{1/2} = \tilde{\nu}_{\sigma_h} \mod 4
\end{eqnarray}
When $|\mu|>{\mu_c}$, 
$|w_{\rm 1D}^{j_z}|=1$ for each $j_z$. This indicates
that there must be {four} zero modes at $k_x=k_y=0$ in each mirror
subsector, which explains why the surface states {in one mirror sector} in
Figs.~\ref{fig:mu1.25} pass the high symmetric points $k_y=0$ or $k_{1\bar{1}0}=0$ when
$\mu=-1.25m_0$.

\subsection{Topological phase diagram}
We summarize the obtained topological phase diagram in Fig.~\ref{fig:TPdiag}.
On the bottom of Fig.~\ref{fig:TPdiag} with $|\mu|=0$, 
 the topological indices and the corresponding surface states originate from those in the normal state. 
As already discussed in Sec.~\ref{sec:normal}, octahedral or cubic Dirac points appear
at topological phase transition points in this regime.
In contrast, for $|\mu|>\mu_c$, as shown in Fig.~\ref{fig:TPdiag}(f) [(e) and (g)],
eight (six) point nodes of the superconducting gap appear on the outer Fermi surface at $v_2=v_1/2$ ($3v_1$ and $-v_1/3$).
These point nodes are also caused by the topological phase {transitions} where the bulk topological number $w_{\mathrm{3D}}$ changes by eight (six). 
These kinds of nodal structures have been overlooked in the group theoretical classification of gap nodes~\cite{sigrist1991},
but our topological analysis clarifies that the presence of the point nodes is not accidental
at the same level as the bulk Dirac points in the normal states.
The presence of point dips near the topological phase transitions provides a chance to confirm the $A_{1u}$ state through low-temperature measurements of
the specific heat, nuclear magnetic resonance spectrum, etc. 

In Fig.\ref{fig:TPdiag}, we also illustrate patterns of helical Majorana fermions composed of electrons of the {outer and inner} Fermi surfaces.
The plus and minus signs in Fig.\ref{fig:TPdiag} denote the helicities of the Majorana fermions, where 
the total helicities for the outer (inner) Fermi surface should be the same as the winding number $w_{\rm out}$ ($w_{\rm in}$).
As explained below, these configurations are determined by comparing the surface state spectra in Fig.\ref{fig:mu1.25} with the winding numbers and considering the requirement from symmetry.  

Let us first consider the region $-1/3<v_2/v_1<1/2$, which includes the spherical symmetric point at $v_2=0$. 
See Fig.~\ref{fig:mu1.25}(b) and Fig.\ref{fig:TPdiag}(b). 
In this region, the outer Fermi surface with $w_{\mathrm{out}}=1$ 
provides a single helical Majorana fermion at the center of the surface Brillouin zone, 
which we denote as the blue symbols and cones in Figs.~\ref{fig:mu1.25}(b) and \ref{fig:TPdiag}(b), respectively, at $k=0$ ($k=k_x$ or $k_{1\bar{1}1}$).  
The inner Fermi surface has an unusual higher winding number 
$w_{{\mathrm{in}}} = 3$.
In the spherical symmetric limit $v_2=0$, we have a single surface state at the center of the surface Brillouin zone, but the 
surface spectrum shows a higher-order $k$ dependence near $k=0$.  
Moreover, when $v_2\neq 0$, the surface state breaks into several helical Majorana fermions because the spherical symmetry reduces to the point group $O_h$. 
For consistency with $C_{4v}$ symmetry on the surface,
there should be at least a single helical Majorana  fermion at the center of the surface Brillouin zone and four others at other points, as illustrated in Fig.\ref{fig:TPdiag} (b). 
In terms of the winding number, this splitting corresponds to $w_{\mathrm{in}} = - 1+4$. 
From the actual surface spectrum in Fig.\ref{fig:mu1.25}(b), 
it is observed that the four Majorana fermions are located on the diagonal mirror lines in the surface Brillouin zone.
We also observe that the two center helical Majorana fermions on the inner and outer Fermi surfaces do not mix, although they are located at the same position with {opposite} helicities.
This is because they also have different topological numbers, $w_{\rm 1D}^{1/2}=w_{\rm 1D}^{3/2}=1$.

To generate the center and satellite Majorana fermions simultaneously,
the surface state has non-monotonic dispersion. We can observe this behavior in Fig.~\ref{fig:mu1.25}(b) for example, 
where a surface state connecting the bulk bands from $k_{1\bar{1}0}\sim -0.7k_0$ to $0.7k_0$ cuts 
the zero energy upward in $k_{1\bar{1}0}<0$, downward at $k_{1\bar{1}0}=0$, and upward again in $k_{1\bar{1}0}>0$. 
This non-monotonic behavior enhances the density of states around $E=0$,
providing a characteristic zero bias conductance peak in 
the tunnel conductance spectra~\cite{yamakage2012}.

We can also determine the configurations of surface Majorana fermions in other regions.
Similar to the argument for the bulk winding numbers in Sec.~\ref{sec:W3D}, we use the duality relation to determine the configuration in the region $1/2<v_2/v_1<3$. 
As the duality retains the spectrum but reverses the winding numbers, the configuration in Fig.\ref{fig:TPdiag} (c) is obtained from Fig. \ref{fig:TPdiag}(b).
Finally, the configurations in Figs.\ref{fig:TPdiag}(a) and (d) are determined by considering the topological phase transitions at $v_2/v_1=-1/3$ and $3$.
At these transition points, six point nodes appear on the outer Fermi surface.  
Upon being projected on the surface Brillouin zone, these point nodes change the helicity of the center Majorana fermions by two, and create four additional {satellite} Majorana fermions in the outer Fermi surface excitations, as shown in Figs.\ref{fig:TPdiag} (a) and (d).
The obtained Majorana configurations in Figs.\ref{fig:TPdiag} (a), (c), and (d) are consistent with the spectra in Figs.\ref{fig:mu1.25} (a), (c) and (d), respectively.

\section{Discussion}\label{sec:discussions}

\subsection{Application to Sr$_{3-x}$SnO}
Notably, a class of antiperovskite materials A$_3$BX with 
A=(Ca, Sr, La), B=(Pb. Sn), and X=(C, N, O) realizes a higher-spin
system. Near the Fermi level, they host a $d$-orbital band of the A atom and
a $p$-orbital band of the B, which belong to the $G_{3/2g}$ and
$G_{3/2u}$ representations of the $O_h$ group, respectively.  
Therefore, our theory is directly applicable to antiperovskite
materials.~\footnote{In antiperovskites, the opposite parities of two $J=3/2$ states
originate from those of the atomic orbitals of A and B. 
However, in general, not only the atomic orbital degrees of freedom 
but also sublattice ones can form the two states with different parities discussed here~\cite{Cano2018}. 
In fact, in the spin $1/2$ TIs of the Bi$_2$Se$_3$ family, 
orbital states with odd and even parity both originate from $p$-orbitals forming 
bonding and anti-bonding states between different sublattices.
The two $J=3/2$ states with opposite parity can also be obtained through this mechanism in general.}

The first-principle calculations for Sr$_3$SnO 
were carried out based on different packages, VASP~\cite{hsieh2014}, WIEN2k, and AkaiKKR~\cite{ikeda2018}. 
These results have scheme dependence in the detailed band structure, such as behavior near the Brillouin zone boundary. 
However, at least, the presence of band inversion at the $\Gamma$ point is consistent in all the schemes.

The first-principle calculations show that the antiperovskites support
the {octahedral} Dirac points with a tiny gap in the $j_z=3/2$ sector~\cite{kariyado2011, kariyado2012, hsieh2014}. 
Therefore, they are near the topological {phase} transition at
$v_2=3v_1$~\cite{hsieh2014}.
In addition, in the antiperovskite with valency (A$^{+2}$)$_3$(B$^{-4}$)(X$^{-2}$), 
$p$-orbitals of B$^{-4}$ ions are almost filled but $d$-orbitals of A$^{+2}$ are almost empty~\cite{kariyado2011,kariyado2012,hsieh2014,oudah2016}.
Therefore, in the antiperovskites with stoichiometric ratio, the Dirac point is pinned near the Fermi level, 
corresponding to $\mu=0$ in our model Hamiltonian in Eq.~(\ref{eq:normal}).

Recently, the superconductivity of antiperovskites has been experimentally observed in Sr$_{3-x}$SnO with the deficiency $x$~\cite{oudah2016, hausmann2017, oudah2018}.
This deficiency yields hole doping. The Fermi level is 
expected to be much lower than the fourfold degenerate point at $\Gamma$.
Therefore, this system can realize the high winding superconductivity in $|\mu|>\mu_c$ discussed in the present work 
if the interorbital attractive interaction is dominant.

Finally, we discuss possible experimental signals of high winding topological superconductivity in the case of Sr$_{3-x}$SnO. 
As mentioned before, Sr$_{3-x}$SnO corresponds to our model with $v_2/v_1\sim 3$ and $\mu<-\mu_c$. 
This indicates that 
if the system realizes the $J=0$ odd-parity superconductivity, 
Sr$_{3-x}$SnO is in the vicinity of the topological phase transition 
between the $w_{\mathrm{out}}=-1$ phase and the $w_{\mathrm{out}}=5$ one 
in our topological phase diagram in Fig.~\ref{fig:TPdiag}. 
Therefore, as shown in Figs~\ref{fig:gapstructure} (a)-(c), there should be point dips on the outer Fermi surface. 
Note that there is no gap node on the inner Fermi surface, 
as the topological phase transition retains the value of $w_{\mathrm{in}}$.
The point node structure can be detected via low-temperature 
experiments such as heat capacity measurement.
Notably, the difference in behaviors of the superconducting gap 
between the outer and inner Fermi surfaces might explain the double superconducting transition 
reported in Sr$_{3-x}$SnO~\cite{oudah2016}.

\subsection{Systems without band inversion}\label{sec:nobandinv}
In the present work, for discussing the topological property of normal and $A_{1u}$ superconducting states, 
we have assumed $m_0\alpha>0$ where the system exhibits a band inversion of $G_{3/2g}$ and the $G_{3/2u}$ representation at the $\Gamma$ point.
In contrast, when $m_0\alpha<0$, without band inversion, the normal state is topologically trivial with the mirror Chern number 
$\nu_h=\nu_d=0$ in any parameter region of $v_1$ and $v_2$.

\begin{figure}[bt]
\includegraphics[width=85mm]{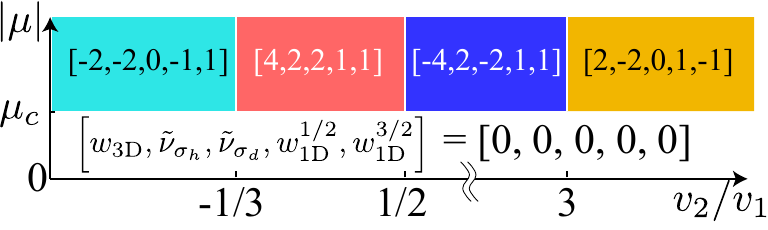}
\caption{Topological phase diagram of the $A_{1u}$ pairing state for $m_0\alpha<0$.
The other conditions are the same as in Fig.~\ref{fig:TPdiag}.
}
\label{fig:TPdiag_nm0}
\end{figure}

For the $A_{1u}$ superconducting state in this case, 
the low-doped phase $|\mu|<\mu_c$ is also topologically trivial as it can be derived in a manner similar to Sec~\ref{sec:mCh}.
However, in the case without band inversion, the topological phase transition 
by doping, which shrinks the Fermi surface to one point, occurs at $|\mu|=\mu_c$.
When the system changes from the low-doped phase $|\mu|<\mu_c$ to the high-doped one $|\mu|>\mu_c$, 
the mirror Chern number jumps with the opposite sign to the case with $m_0\alpha>0$ 
as the angular momenta of valence and conduction bands at the $\Gamma$ point (see Fig.~\ref{fig:level} for $m_0\alpha>0$)
exhibit opposite signs for the cases with and without band inversion.
In addition, applying the analysis in Sec.~\ref{sec:W3D} and \ref{sec:W1D}
we can derive the 3D and 1D winding numbers with the same value as $m_0\alpha>0$. 
Consequently, all the topological indices for the high-doped phase of $m_0\alpha<0$ 
are the same as those in $m_0\alpha>0$, as summarized in Fig.~\ref{fig:TPdiag_nm0}.

In contrast, when $m_0\alpha<0$, without band inversion, 
the orbital mixing in both the Fermi surfaces is strongly suppressed.
This reduces the transition temperature of the interorbital pairing state $A_{1u}$. 
Thus, although the topological property of the high-doped phase is unchanged regardless of the sign of $m_0\alpha$, 
the band inversion plays a key role in obtaining the new topological superconductivity 
discussed in this work. 

\subsection{Other pairing states}\label{sec:pairing}
Depending on the pairing interaction, we cannot avoid the possibility of the $s$-wave superconductivity 
belonging to the $A_{1g}$ representation in Table~\ref{table:Oh}. 
As discussed in Sec.~\ref{sec:Tc}, it exhibits the highest transition temperature in even-parity superconductivity.
In this case, not only the 3D winding number, but also the symmetry-protected 1D winding numbers and mirror Chern numbers
in the superconducting state are trivial.

However, the $s$-wave superconducting state may host surface topological superconductivity. 
Owing to the non-zero mirror Chern numbers in the normal state, 
the system supports surface Dirac fermions in the normal state. 
In the $s$-wave superconducting state, the surface Dirac fermion 
exhibits surface topological superconductivity~\cite{fang2014}.

The $A_{1u}$ and $A_{1g}$ states with $J=0$ mainly discussed so far 
are caused by the Cooper pair between time-reversal partners.
However, this is not the only possible partner.  
An electron may also 
form a Cooper pair with its inversion partner. 
Such a Cooper pair inevitably has a parallel spin configuration with
$J=1, 3$ as inversion does not flip the spin.
As shown in Table \ref{table:Oh}, 
the $A_{2u}$, $T_{1u}$, and $T_{2u}$ states contain this class of Cooper pairs. 

As shown in Fig.~\ref{fig:Tc}, $T_{1u}$ exhibits a relatively higher $T_{\mathrm{c}}$. 
Notably, $T_{\mathrm{c}}$ of $T_{1u}$ is comparable to that of
$A_{1u}$, which suggests that $T_{1u}$ could win by improving the spin and orbital 
textures slightly. 
Indeed, a similar exchange of the pairing symmetry has been reported in the case of doped TIs with spin $1/2$. 
The higher-order term with respect to $k$ enhances the $T_{\mathrm{c}}$ of 
the spin-triplet $E_{u}$ representation of $D_{3d}$ symmetry more than that of the spin-singlet $A_{1u}$ one~\cite{Fu2014}.
It supports the experimental observation of the anisotropic gap~\cite{Matano2016,Yonezawa2016}.
Similarly, the present calculation does not exclude the possibility of $T_{1u}$.

Let us consider the gap structure and spin property of the $T_{1u}$ state. 
It hosts $J=1$ and $J=3$ components, 
and hence, the gap function is given as 
\begin{eqnarray}\label{eq:t1u}
	\Delta = \frac{2\Delta_0}{\sqrt{5}}(\chi_1 J_z + \chi_3 \tilde{J}_z)\sigma_y
\end{eqnarray}
with the normalization condition of the mixing ratio $\chi_1^2 + \chi_3^2 = 1$.
Projecting Eq.~(\ref{eq:t1u}) onto the band basis of the normal state,
we obtain the energy gap at the Fermi surface of $n$-th band as [see Appendix~\ref{sec:PTSC}]
\begin{eqnarray}\label{eq:invproj}
	d'_{n}(\bm{k}) &&= -\langle u_{n,{\bm k}}| \Delta | u_{\bar n,{\bm k}}\rangle \nn\\
	&&= \chi_1 d^{\prime(1)}_n(\bm{k}) + \chi_3 d^{\prime(3)}_n(\bm{k}), 
\end{eqnarray}
where we use the fact that the Cooper pair of the $T_{1u}$ state is formed by the electron and its inversion partner. 

\begin{figure}[t]
\includegraphics[width=85mm]{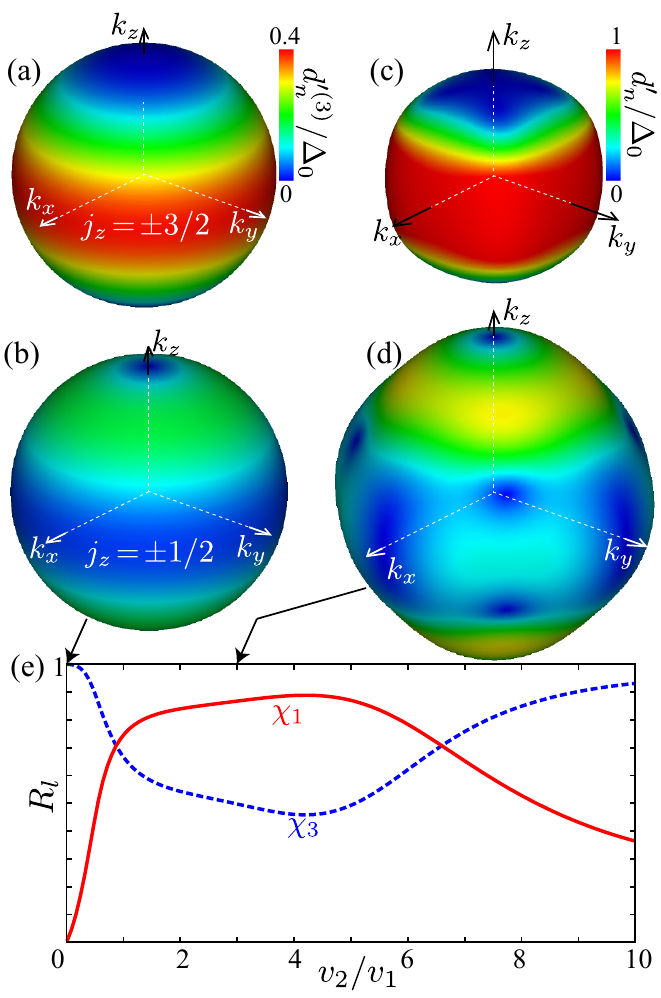}
\caption{Energy gap $d'_{n}$ of $T_{1u}$ representation on the Fermi surfaces for $v_{2}=0$ (a,b) and $3v_1$ (c,d). 
(a,c) and (b,d) show the inner and outer Fermi surfaces, respectively.
(e) Mixing ratio of angular momenta $l=1$ and $l=3$ pairs for $T_{1u}$ gap function $\Delta = \Delta_0\frac{2}{\sqrt{5}}(\chi_1 J_z + \chi_3 \tilde{J}_z)\otimes \sigma_y$.
The parameters are the same as in Fig.~\ref{fig:Tc}.
}
\label{fig:mixing_T1u}
\end{figure}

Along the main axis $k_z$, the eigenvalue $\lambda_{C_4}=e^{-i \frac{j_z}{2}\pi}$ of $C_4$ rotation is a good quantum number of the normal state. 
The band $n$ and $\bar{n}$ in Eq.~(\ref{eq:invproj}) belong to a different eigensector with $\lambda_{C_4}$ and $-\lambda_{C_4}$. 
In addition, as $J_z$ and $\tilde{J}_z$ preserve $\lambda_{C_4}$, 
there is a point node $d'_{n}(k_z)=0$ at the intersection of the $k_z$ axis and the two Fermi surfaces. 
The presence of the point node is consistent with the group theoretical classification of the gap nodes~\cite{sigrist1991}.
Along the other momentum direction, adjusting 
the ratio of these two components $\chi_1$ and $\chi_2$, the 
$T_{1u}$ state can optimize the gap function so as to be consistent with the spin and
orbital textures on the Fermi surfaces, as much as possible.
We illustrate the optimized superconducting gap of the $T_{1u}$ state for $v_2=3 v_1$ ($v_2=0$)
in Figs.~\ref{fig:mixing_T1u} (c) and (d) (Fig.~\ref{fig:mixing_T1u} (a) and (b)).

In Fig.~\ref{fig:mixing_T1u}(e), we show the parameter dependence of the optimum mixing ratio
of the $J=1$ and $J=3$ components at $T_{\mathrm{c}}$ in our numerical solution of the gap
equation. It is observed that the higher-spin $J=3$ component cannot be
neglected in the whole region of $v_2/v_1$, 
and it can even dominate the gap function in some regions.
Therefore, the $T_{1u}$ state realizes a higher-spin pairing state.
The higher-spin nature of the pairing state could be tested via spin-
sensitive experiments.

In particular, at the rotational symmetric point $v_2=0$, 
the amplitude of the total momentum is a good quantum number.
Thus, the $J=1$ state and $J=3$ state do not mix with each other. 
Let us compare the stability of these two superconductivities in terms of the gap structure.
From the analytical form of wave function (\ref{eq:wfnsph}), the
$J=1$ component is given as 
\begin{eqnarray}
	&&d_{j_z,\pm}^{\prime(1)}(\hat{\bm k}) = \pm \frac{2\Delta_0}{\sqrt{5}}\cos(2\rho_{k}) 
	\langle j_z| \bm{J}\cdot\hat{\bm k} | -\!j_z \rangle,  
\end{eqnarray}
for two Fermi surfaces with the angular momentum $j_z=\pm 3/2$ and $\pm 1/2$ along $\bm{k}$.
Here, as the spin operators $J_x$ and $J_y$ shift $j_z$ only by $\pm 1$ and $J_z$ preserves $j_z$, 
the $J=1$ component cannot couple the $j_z\!=\!\pm \frac{3}{2}$ states, i.e., 
\begin{eqnarray}
	d_{\pm{3}/{2},\pm}^{\prime(1)}(\hat{\bm k})=0
\end{eqnarray}
for all the directions of $\bm{k}$. 
In contrast, the $J=3$ component opens the energy gap in this Fermi surface 
because $\tilde{J}_x$ and $\tilde{J}_y$ have finite matrix elements between $j_z= \pm\frac{3}{2}$ states.
[see Appendix~\ref{sec:Jmat}] 
Therefore, it is observed that the $J=3$ state is more stable than $J=1$ at this point.

Notably, the $T_{1u}$ state realizes a nematic
superconducting state, which spontaneously {breaks} the $O_h$ group symmetry. 
As illustrated in Fig.~\ref{fig:mixing_T1u}(a-d), $T_{1u}$ supports an
anisotropic superconducting gap breaking the $O_h$ group.
The nematic feature can also be used to identify the $T_{1u}$ state
through thermal transport measurements~\cite{Yonezawa2016}.

\section{Conclusions}\label{sec:conclusions}
We have investigated topological superconductivity in
doped TIs of $J=3/2$ electrons. 
Antiperovskites realize this class of topological materials, whose normal state hosts Dirac points at the topological quantum phase
transition.
In contrast to half-Heuslers with a single $J=3/2$ band, the antiperovskites support
multiple $J=3/2$ bands, which allows us to realize
various unconventional superconductivities within the BCS
type of constant gap functions.  
Using the $k\cdot p$ Hamiltonian, 
we {have demonstrated} that the $J=0$ odd-parity pairing state
{with the $A_{1u}$ representation of $O_h$ group} exhibits the highest transition temperature $T_{\mathrm{c}}$ if the
interorbital pairing interaction is dominant.
The $A_{1u}$ state shows {a new class of topological superconductivity}
with a higher winding number $w_{3D}=\pm2$ {or} $\pm4$.
Moreover, it displays topological crystalline superconductivity
with respect to the mirror reflections and fourfold rotation of the
$O_h$ group. 

We have also revealed that the nematic $T_{1u}$
state with $J=3$ Cooper pairs has a high $T_{\mathrm{c}}$, which is 
comparable to that of $A_{1u}$. 
This result suggests an interesting possibility of spin-septet
nematic superconductivity in antiperovskite materials.  
Similar to the case of doped Bi$_2$Se$_3$, 
the $T_{1u}$ phase might be stabilized against the $A_{1u}$ phase
by considering the higher-order terms of the $k\cdot p$ Hamiltonian.

\begin{acknowledgments}
This work was supported by JSPS KAKENHI Grant numbers JP16K17755, JP16H06861, JP17H02922, JP17J08855, and
the JSPS Core-to-Core program. 
S.K. is also supported by the Building of Consortia for the Development of Human Resources in Science and Technology.
\end{acknowledgments}

\appendix
\section{Spin matrices in $J=3/2$ space}\label{sec:Jmat}
Here, we provide the explicit forms of the matrices useful to describe $J=3/2$ systems.
First, the spin matrices in the $J=3/2$ space are given by 
\begin{eqnarray}
J_x=
\frac{1}{2}\left(\begin{array}{cccc}
0 & \sqrt{3} &0 & 0 \\
\sqrt{3} & 0 & 2 & 0 \\
0 & 2 & 0 & \sqrt{3} \\
0 & 0 & \sqrt{3} & 0 
\end{array}\right), \\
J_y=
\frac{i}{2}\left(\begin{array}{cccc}
0 & -\sqrt{3} &0 & 0 \\
\sqrt{3} & 0 & -2 & 0 \\
0 & 2 & 0 & -\sqrt{3} \\
0 & 0 & \sqrt{3} & 0 
\end{array}\right), \\
J_z=
\frac{1}{2}\left(\begin{array}{cccc}
3 & 0 &0 & 0 \\
0 & 1 & 0 & 0 \\
0 & 0 & -1 & 0 \\
0 & 0 & 0 & -3 
\end{array}\right).
\end{eqnarray}

In general, the $4\times 4$ Hermitian matrix in the $J=3/2$ space is expanded by 
the basis given in Table~\ref{table:Sclass}. 
Among them , the $T_{1u}$ representation including spin matrices itself $\{J_x, J_y, J_z\}$ 
behaves as a vector under the spin rotation $C_{q \bm{n}}$ defined in Eq.~(\ref{eq:Cqn}). 
Here, the other set of $T_{1u}$ basis is given by the third-order polynomials of spin matrix, 
which has an explicit form as follows:
\begin{eqnarray}
\tilde{J}_x &\equiv& \frac{5}{3}(J_yJ_xJ_y+J_zJ_xJ_z) - \frac{7}{6}J_x \nn\\
&=&\frac{1}{4}\left(\begin{array}{cccc}
0 &  \sqrt{3} &0 & -5 \\
\sqrt{3} & 0 & -3 & 0 \\
0 & -3 & 0 &  \sqrt{3} \\
-5 & 0 & \sqrt{3} & 0 
\end{array}\right), \\
\tilde{J}_y &\equiv& \frac{5}{3}(J_zJ_yJ_z+J_xJ_yJ_x) - \frac{7}{6}J_y \nn\\
&=&\frac{i}{4}\left(\begin{array}{cccc}
0 & -\sqrt{3} &0 & -5 \\
\sqrt{3} & 0 & 3 & 0 \\
0 & -3 & 0 & -\sqrt{3} \\
5 & 0 & \sqrt{3} & 0 
\end{array}\right), \\
\tilde{J}_z &\equiv& \frac{5}{3}(J_xJ_zJ_x+J_yJ_zJ_y) - \frac{7}{6}J_z \nn\\
&=&\frac{1}{2}\left(\begin{array}{cccc}
-1 & 0 &0 & 0 \\
0 & 3 & 0 & 0 \\
0 & 0 & -3 & 0 \\
0 & 0 & 0 & 1 
\end{array}\right).
\end{eqnarray}
By this definition, the two sets of $T_{1u}$ basis satisfy the orthogonal condition 
\begin{eqnarray}
	\bm{J}\cdot\bm{J} = \tilde{\bm J}\cdot\tilde{\bm J} = \frac{15}{4}, \quad {\bm J}\cdot\tilde{\bm J}=0.
\end{eqnarray}

\section{Superconductivity with time-reversal and inversion symmetries}\label{sec:PTSC}
\subsection{Nambu space and symmetry}\label{sec:Nmbsym}
In this section, we summarize the general properties of the BdG Hamiltonian
for the systems with time-reversal symmetry.
Here, we use the Nambu space spanned by the basis 
$(\bm{c}_{\bm{k}},\ \bar{\bm{c}}_{-\bm{k}})$, 
where the spinor $\bm{c}_{\bm{k}}$ consists of the annihilation operator $c_{a, \bm{k}}$
with the spin and orbital indices $a= (\sigma_z, j_z)$, and 
$\bar{\bm{c}}_{-\bm{k}}$ is its time-reversal hole partner with the components
$\bar{c}_{a,-\bm{k}}=\sum_{a'}(C_{2,\bm{y}})_{aa'}{c}_{a',-\bm{k}}^\dag$.
In this basis, the BdG Hamiltonian is written as 
\begin{eqnarray}\label{eq:bdg_app}
	H(\bm{k}) = 
	\left(\begin{array}{cc} 
	H_0(\bm{k}) & \Delta(\bm{k}) \\
	\Delta^\dag(\bm{k})& -H_0(\bm{k}) \\
	\end{array}\right).
\end{eqnarray} 
where we have used the time-reversal symmetry of the one-particle Hamiltonian in the normal state (\ref{eq:TRS_N}). 

The point group operation for the particle space is given as 
${c}_{a \bm{k}} \rightarrow \sum_a G_{aa'}c_{a', \bm{k}}$
with $G=C_{q,\bm{n}}$, $P$, or $PC_{q,\bm{n}}$ 
corresponding to discrete rotation, inversion, and their combined operation 
acting on the internal degrees of freedom $a=(\sigma_z,j_z)$.
 The transformation $G$ for the time-reversal hole is 
$\bar{c}_{a,-\bm{k}}\rightarrow \sum_{a',b'}(C_{2,\bm{y}})_{aa'} (\mathcal{K} G_{a'b'}\mathcal{K}^{-1}) {c}^\dag_{b',-\bm{k}}  
=\sum_{a',b'} G_{a, a'}(C_{2,\bm{y}})_{a'b'} {c}^\dag_{b',-\bm{k}} = \sum_{a'} G_{aa'}\bar{c}_{a',-\bm{k}}$,
because the time-reversal operator ${\cal T}=C_{2,\bm{y}}\mathcal{K}$ commutes with any point group operation $G$.
The transformations for the particle and time-reversal hole components are similar. 
$\mathcal{T}$ and $G$ act on the gap function as 
\begin{eqnarray}
	\mathcal{T}\Delta(-\bm{k})\mathcal{T}^{-1} \ \hbox{and} \ G\Delta(D_{G}^{-1}[\bm{k}])G, 
\end{eqnarray}
where $D_{G}[\bm{k}]$ indicates the $G$ operation of $\bm{k}$. When $\mathcal{T}\Delta(-\bm{k})\mathcal{T}^{-1}=\Delta(\bm{k})$, 
the BdG Hamiltonian is time reversal invariant,
\begin{eqnarray}
	&&\tilde{\mathcal{T}} H(\bm{k}) \tilde{\mathcal{T}}^{-1} = H(-\bm{k}) \quad\hbox{with}\quad \tilde{\mathcal{T}} = \mathcal{T}\tau_0 \label{eq:TRSbdg}
\end{eqnarray}


The Fermi statistics of $\bm{c}_{\bm k}$ result in a restriction on the gap function  
\begin{eqnarray}\label{eq:parityconst}
	(\Delta({\bm k}) C_{2,\bm{y}})^T =-\Delta(-{\bm k}) C_{2,\bm{y}}.
\end{eqnarray}
To satisfy both (\ref{eq:TRSbdg}) and (\ref{eq:parityconst}), the gap function is Hermitian:
\begin{eqnarray}\label{eq:delher}
	\Delta^\dag(\bm{k}) = \Delta(\bm{k}). 
\end{eqnarray}
Consequently, the BdG Hamiltonian (\ref{eq:bdg_app}) for the time-reversal symmetric superconductivity is reduced to
\begin{eqnarray}\label{eq:bdgtauzx}
	H(\bm{k}) = H_0(\bm{k}) \tau_z + \Delta(\bm{k}) \tau_x, 
\end{eqnarray}
where $\tau_i$ represents the Pauli matrices acting on the particle and time-reversal hole space.

We also consider the particle-hole symmetry of the BdG Hamiltonian 
\begin{eqnarray}\label{eq:phs}
	\mathcal{C}H(\bm{k})\mathcal{C}^{-1}\!=\!-\!H(-\bm{k}) \quad \hbox{with}\quad
	\mathcal{C}=
	\left(\begin{array}{cc}
		0 & C_{2,\bm{y}}^\dag \\
		C_{2,\bm{y}} & 0 \\
	\end{array}\right)\mathcal{K}, \nn\\
\end{eqnarray}
inherent to superconductors. 
By combining (\ref{eq:phs}) and (\ref{eq:TRSbdg}), chiral symmetry is given as
\begin{eqnarray}\label{eq:chs}
 \Gamma H(\bm{k}) \Gamma=-H(\bm{k}) \quad \hbox{with}\quad \Gamma=-i\tilde{\mathcal{T}}\mathcal{C}=\tau_y.
\end{eqnarray}

\subsection{Band representation}\label{sec:bandrep}
In this section, we consider the BdG Hamiltonian in terms of the band basis of the normal state.
This basis is given by the Bloch equation of the normal state
\begin{eqnarray}\label{eq:schrodinger}
	H_0(\bm{k}) |u_{n,\bm{k}} \rangle = \xi_n(\bm{k}) |u_{n,\bm{k}} \rangle.
\end{eqnarray}
with the band index $n$ and eigenenergy $\xi_{n}(\bm{k})$. 
From orthonormality and completeness of basis, 
\begin{eqnarray}\label{eq:comp}
	\langle {u}_{n, k} | {u}_{n', k}\rangle = \delta_{n n'}, \quad \sum_{n} |{u}_{n, k}\rangle \langle {u}_{n, k}| = \hat{1},
\end{eqnarray}
the one-particle Hamiltonian is written as a diagonal form on this basis
\begin{eqnarray}\label{eq:h0band}
	H_0(\bm{k}) = \sum_{n}\xi_{n}(\bm{k}) |{u}_{n,\bm{k}}\rangle \langle {u}_{n,\bm{k}}|.
\end{eqnarray}

For later convenience, we consider the time-reversal and inversion symmetry of the basis. 
From~(\ref{eq:TRS_N}) for the one-particle Hamiltonian of the normal state, 
the time-reversal partner of $|u_{n,\bm{k}} \rangle$ is 
\begin{eqnarray}
	|u_{\bar n,-\bm{k}} \rangle = \mathcal{T} |u_{n,\bm{k}} \rangle,
\end{eqnarray}
with $\bar{n}\neq n$ because of $\mathcal{T}^2=-1$. 
From (\ref{eq:inv}), 
the inversion partner is 
\begin{eqnarray}\label{eq:Pwfn}
	|u_{n,-\bm{k}} \rangle = P |u_{n,\bm{k}} \rangle.
\end{eqnarray}
In the presence of both symmetries, the energy of the state
\begin{eqnarray}\label{eq:Twfn}
	|u_{\bar{n},\bm{k}} \rangle = P\mathcal{T} |u_{n,\bm{k}} \rangle
\end{eqnarray}
is degenerated with that of $|u_{n,\bm{k}} \rangle$, i.e., $\xi_{\bar{n}}=\xi_{n}$. 

Let us consider the band representation of the gap function. 
As both $|u_{n, \bm{k}}\rangle$ and $|u_{\bar{n}, \bm{k}}\rangle$ have completeness (\ref{eq:comp}), 
we can expand the gap function as
\begin{eqnarray}\label{eq:delp}
	\Delta(\bm{k}) 
	= -\sum_{n, n'}d_{n,n'}(\bm{k}) |{u}_{n,\bm{k}} \rangle \langle {u}_{\bar n',\bm{k}}|
\end{eqnarray}
with the matrix element
$d_{n,n'}=-\langle{u}_{n,\bm{k}}| \Delta(\bm{k}) |{u}_{\bar n',\bm{k}}\rangle = -\langle{u}_{n,\bm{k}}| \Delta \mathcal{T}|{u}_{n',-\bm{k}}\rangle$. 
The $d_{n,n'}$ provides the superconducting correlation
for Cooper pairs consisting of $|u_{n,\bm{k}}\rangle$ and $|u_{n',-\bm{k}}\rangle$. 
In the weak coupling limit where the gap function $\Delta$ is much smaller than the level 
spacing among the different bands, 
the matrix element $d_{n,n'}$ is non-zero 
only for the degenerated bands $n'=n$ or $n'=\bar{n}$ [see Eq.(\ref{eq:Twfn})]. 
This indicates that the time-reversal or inversion partners can form the Cooper pair.
These pairing states are described by the matrix elements 
\begin{eqnarray}\label{eq:dweak}
	d_{n,n'}=
	\left\{\begin{array}{l}
		d_n \delta_{n,\bar{n}'} \hbox{ for time-reversal pair} \\
		d_n' \delta_{n,n'} \hbox{ for inversion pair} 
	\end{array}\right. .
\end{eqnarray}

In particular, when the time-reversal partners form the Cooper pair,
not only the normal Hamiltonian (\ref{eq:h0band})  
but also the gap function is diagonal in the band representation [see Eq.~(\ref{eq:dweak})]. 
Therefore, the BdG Hamiltonian is also diagonal
\begin{eqnarray}\label{eq:bdgdiag_tauzx}
	H(\bm{k})\! =\! \sum_{n}\Big(\xi_n(\bm{k})\tau_z\! -\! d_n(\bm{k})\tau_x \Big) |u_{n,\bm{k}}\rangle\langle u_{n,\bm{k}}|, 
\end{eqnarray}
where $d_n=\langle u_{n,\bm{k}}| \Delta |u_{n,\bm{k}} \rangle$ is real owing to hermiticity (\ref{eq:delher}). 
Note that the time-reversal and inversion symmetries for the gap function are
\begin{eqnarray}\label{eq:trsinvdel}
	d_{\bar n}(-\bm{k}) = d_{n}(\bm{k}) \\
	d_{n}(-\bm{k}) = \pm d_{n}(\bm{k}), 
\end{eqnarray}
where the double sign indicates even- and odd-parity superconductivities. 

\subsection{3D winding number for time-reversal pairs}
The time-reversal and particle-hole symmetries defined by Eqs.~(\ref{eq:TRSbdg}) and (\ref{eq:phs}) 
do not depend on any particular crystal structure. 
They thus specify the most general symmetry-protected topological number~\cite{schnyder2008}.
In the present system with $\mathcal{T}^2=-1$ and $\mathcal{C}^2=+1$,
the general topological number is the 3D winding number
\begin{eqnarray}\label{eq:w3ddef}
	w_{\mathrm{3D}} \!=\!\! \int\!\! \frac{d^3k}{48\pi^2} 
	\epsilon^{\alpha\beta\gamma} 
	\mathrm{Tr}\!
	\left[\Gamma
	(H^{-1} \partial_\alpha H)
	(H^{-1} \partial_\beta H)
	(H^{-1} \partial_\gamma H)
	\right].\nn\\
\end{eqnarray}
with the 3D Levi-Civita symbol $\epsilon^{\alpha\beta\gamma}$. 
For the time-reversal pairing state, the diagonal form of the Hamiltonian (\ref{eq:bdgdiag_tauzx}) provides a simple analytic form of $w_{\mathrm{3D}}$. 
Here, we derive it by applying the procedure in Ref.~\cite{qi2010} 
to the system with the additional inversion symmetry. 

First, we consider the unitary transformation $(\tau_x,\tau_y,\tau_z)\rightarrow (\tau_y,\tau_z,\tau_x)$. 
It yields the eigenbasis of the chiral operator with $\Gamma=\tau_z$, convenient to analyze $w_{\mathrm{3D}}$.
In this basis, the BdG Hamiltonian (\ref{eq:bdgdiag_tauzx}) is written as
\begin{eqnarray}\label{eq:bdgoffdiag}
	H(\bm{k}) \!&\!=\!&\! \sum_{n}(\xi_n(\bm{k})\tau_x - d_n(\bm{k})\tau_y ) 
	|u_{n,\bm{k}}\rangle\langle u_{n,\bm{k}}|\nn\\ 
	\!&\!=\!&\!\left(\begin{array}{cc} 
	0 & h(\bm{k}) \\
	h^\dag(\bm{k})& 0 \\
	\end{array}\right) 
\end{eqnarray}
where the off-diagonal element is 
\begin{eqnarray}
	h(\bm{k})\!=\!\sum_{n} \Big(\xi_n(\bm{k}) + id_n(\bm{k})\Big)|u_{n,\bm{k}}\rangle\langle u_{n,\bm{k}}|.
\end{eqnarray}
\begin{figure}[tb]
\includegraphics[width=85mm]{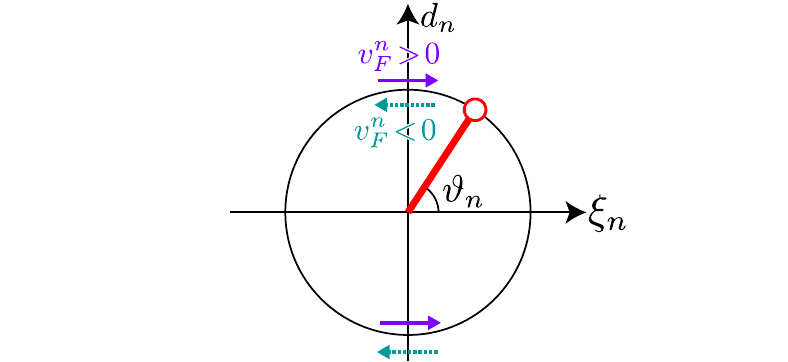}
\caption{Phase factor $\kappa_{n}(k)$ and its path with the increase in $k$.
}
\label{fig:phase}
\end{figure}

Second, we use the so-called spectral flattening technique~\cite{qi2008}.
The eigenenergy of (\ref{eq:bdgoffdiag}) is given by the absolute values of the
off-diagonal elements $E=\pm|\xi_{n}+id_{n}|$.
Therefore, the replacement 
\begin{eqnarray}\label{eq:htoq}
	h(\bm{k})\rightarrow q(\bm{k})&=&\sum_{n} \frac{\xi_{n}(\bm{k})+id_{n}(\bm{k})}{|\xi_{n}(\bm{k})+id_{n}(\bm{k})|}|{u}_{n,\bm{k}} \rangle \langle {u}_{n,\bm{k}}| \nn\\
	                      &=&\sum_{n} e^{i\vartheta_{n}(\bm{k})}|{u}_{n,\bm{k}} \rangle \langle {u}_{n,\bm{k}}|,
\end{eqnarray}
provides the adiabatic transformation of the original Hamiltonian to that with flat bands with $E=\pm1$ without changing $w_{\mathrm{3D}}$.
The phase factor $\vartheta(\bm{k})$ is defined as in Fig.~\ref{fig:phase}.
In this technique, the Hamiltonian is replaced as
\begin{eqnarray}\label{eq:htoq}
	H\rightarrow Q (\bm{k}) = 
	\left(\begin{array}{cc}
	0 & q(\bm{k}) \\
	q^{\dag} (\bm{k}) & 0
	\end{array}\right), 
\end{eqnarray}
and its inverse is replaced as 
\begin{eqnarray}
	H^{-1}\rightarrow Q^{-1} (\bm{k}) =  Q(\bm{k}), 
\end{eqnarray}
where we have used $q^{-1}(\bm{k})=q^{\dag}(\bm{k})$.
Through this replacement, we can evaluate the 3D winding number (\ref{eq:w3ddef}) as
\begin{eqnarray}\label{eq:w3dwf}
	w_{\mathrm{3D}} \!&\!=\!&\! \int \frac{d^3k}{48\pi^2} 
\epsilon^{\alpha\beta\gamma}
\mathrm{Tr}
\left[\Gamma
(Q \partial_\alpha Q)
(Q \partial_\beta Q)
(Q \partial_\gamma Q)
\right] \nn \\
\!&\!=\!&\! \int \frac{d^3k}{24\pi^2} \epsilon^{\alpha\beta\gamma} \mathrm{Tr} \Big[(q\partial_\alpha q^{\dag}) (q\partial_\beta q^{\dag}) (q\partial_\gamma q^{\dag})\Big]. \nn \\ 
\!&\!=\!&\!\int \frac{d^3k}{4\pi^2} \epsilon_{\alpha\beta\gamma} \sum_{n,n'}
	 \Bigg[  (i\partial_{\gamma} \vartheta_{n}) 
	\langle \partial_{\alpha} u_{n,\bm{k}}|\partial_{\beta} u_{n,\bm{k}}\rangle \delta_{n,n'} \nn\\\!&&
	\!-  (i\partial_{\gamma} \vartheta_{n}) \cos \vartheta_{n'n} 
		\langle \partial_{\alpha}u_{n,\bm{k}}| u_{n',\bm{k}} \rangle 
		\langle u_{n',\bm{k}}|\partial_{\beta} u_{n,\bm{k}}\rangle  \nn\\\!&& 
	\!-  i\sin \vartheta_{n'n}	
		\langle  \partial_{\alpha} u_{n',\bm{k}}| \partial_{\beta} u_{n,\bm{k}}\rangle
	    \langle \partial_{\gamma} u_{n,\bm{k}}| u_{n',\bm{k}}\rangle \Bigg] 
\end{eqnarray}
with $\vartheta_{n'n}=\vartheta_{n'}-\vartheta_n$. 
To obtain the last equality in Eq.~(\ref{eq:w3dwf}), 
we have used the time-reversal symmetry for the phase
$\vartheta_{\bar n}(-\bm{k})=\vartheta_{n}(\bm{k})$ given by (\ref{eq:trsinvdel}) 
and for the eigenstates 
$\langle \partial_{\alpha}u_{\bar n,-\bm{k}}|\partial_{\beta}u_{\bar n',-\bm{k}} \rangle=
\langle \partial_{\beta}u_{n',\bm{k}}|\partial_{\alpha}u_{ n,\bm{k}} \rangle$ and 
$\langle \partial_{\alpha}u_{\bar n,-\bm{k}}|u_{\bar n',-\bm{k}} \rangle=
\langle u_{n',\bm{k}}|\partial_{\alpha}u_{ n,\bm{k}} \rangle$
by (\ref{eq:Twfn}) in addition to the usual algebraic expansion. 

Subsequently, we use the weak coupling limit with $d_{n}\rightarrow 0$, 
where the phase factor is given by the step function
\begin{eqnarray}\label{eq:weakcoupling}
	\vartheta_n(\bm{k}) =
	\left\{ \begin{array}{l} 
	0 \hbox{ for } k>k_{\mathrm{F}} \\
	\pi \hbox{ for } k < k_{\mathrm{F}} \\
	\pi/2 \hbox{ for } k = k_{\mathrm{F}} \hbox{ and } d_n>0\\
	-\pi/2 \hbox{ for } k = k_{\mathrm{F}} \hbox{ and } d_n<0
	\end{array}\right. .
\end{eqnarray}
The contribution from the third term in Eq.~(\ref{eq:w3dwf}) is zero in this limit. 
In addition, away from the Fermi surface, the derivative of the phase factor $\partial_\gamma \vartheta_n=0$. 

Let us consider the non-trivial contribution for the 3D winding number in~(\ref{eq:w3dwf}) from momenta close to the Fermi surface. 
Here, we consider the coordinate in momentum space $(k_{\alpha},k_{\beta},k_{\gamma})$ 
as $(k,k'_{1},k'_{2})$ with momenta $k$ and $k'_{\alpha=1,2}$ perpendicular and parallel to 
the Fermi surface, respectively.  The weak coupling limit (\ref{eq:weakcoupling}) yields 
$\partial_{k'_{\alpha}}\vartheta = 0$. In addition, the derivative with respect to $k$ is 
\begin{eqnarray}~\label{eq:phaseW}
	\partial_{k} \vartheta_{n}  = -\mathrm{sign}\Big[v_{\mathrm{F}}^n\Delta_{n} \Big] \pi \delta(k-k_{\mathrm{F}}),
\end{eqnarray}
where $v_{\mathrm{F}}^n$ is the Fermi velocity of band $n$. 
Consequently, the 3D winding number (\ref{eq:w3dwf}) is reduced to 
\begin{eqnarray}\label{eq:w3dred2}
	w_{\mathrm{3D}}  \!&\!=\!&\!-i\mathrm{sign}\Big[v_{\mathrm{F}^n}\Delta_{n} \Big] \sum_{n}\int \frac{d^2k'}{4\pi}  \epsilon_{\alpha\beta} \Big[ 
	\langle \partial_{k_\alpha'} u_{n,\bm{k}}|\partial_{k_\beta'} u_{n,\bm{k}}\rangle \nn\\ \!&&\!
	-\! \sum_{n'}  \cos\vartheta_{n'n}(k_F^n) 
		\langle \partial_{k_\alpha'}u_{n,\bm{k}}| u_{n',\bm{k}} \rangle 
		\langle u_{n',\bm{k}}|\partial_{k_\beta'} u_{n,\bm{k}}\rangle \Big] \nn\\&&
\end{eqnarray}
Equation~(\ref{eq:weakcoupling}) and the time-reversal and inversion symmetries $\vartheta_{\bar n}(\bm{k})=\pm \vartheta_{n}(\bm{k})$ 
yield the phase factor at the Fermi surface as 
\begin{eqnarray}
	\cos (\vartheta_n(k_F^n)-\vartheta_{n'}(k_F^n)) = 
	\left\{ \begin{array}{l}
	1 \hbox{ for }n'= n  \\
	\pm 1 \hbox{ for } n'=\bar n  \\
	0 \hbox{ otherwise } \\
	\end{array}\right.
\end{eqnarray}
where the double sign indicates even- and odd-parity superconductivities.
When $n'=n$, the second term of (\ref{eq:w3dred2}) is zero owing to $\epsilon_{\alpha\beta}
\langle \partial_{k_\alpha'}u_{n,\bm{k}}| u_{n,\bm{k}} \rangle 
\langle u_{n,\bm{k}}|\partial_{k_\beta'} u_{n,\bm{k}}\rangle =0$. 
In addition, even for $n'=\bar{n}$, the second term of (\ref{eq:w3dred2}) is 
\begin{eqnarray}
	&& f_n = \epsilon_{\alpha\beta} 
		\langle \partial_{k_\alpha'}u_{n,\bm{k}}| u_{\bar n,\bm{k}} \rangle 
		\langle u_{\bar n,\bm{k}}|\partial_{k_\beta'} u_{n,\bm{k}}\rangle \nn\\&&
	=
	-\epsilon_{\alpha\beta}
		\langle \partial_{k_\alpha'}u_{\bar n,\bm{k}}| u_{n,\bm{k}}\rangle
		\langle u_{n,\bm{k}}|\partial_{k_\beta'} u_{\bar n,\bm{k}} \rangle
	= -f_{\bar{n}}
\end{eqnarray}
where we have used $\epsilon_{\alpha\beta}=-\epsilon_{\beta\alpha}$ and $\langle\partial_{\alpha}u_{n,\bm{k}}|u_{n',\bm{k}}\rangle = -\langle u_{n,\bm{k}}|\partial_{\alpha}u_{n',\bm{k}}\rangle$. 
Therefore, the contribution from $f_{\bar n}$ cancels that from $f_{n}$ when performing summation over $n$ and $\bar{n}$. 
Consequently, only the first term in (\ref{eq:w3dred2}) yields the non-trivial contribution to $w_{\mathrm{3D}}$. 

Consequently, the 3D winding number is given by
\begin{eqnarray}
	w_{\mathrm{3D}} = \frac{1}{2}\sum_{n} \mathrm{sign}\Big[v_{\mathrm{F}}d_{n} \Big] \nu_{\mathrm{Ch}}^{j_z} 
\end{eqnarray}
with the first Chern number
\begin{eqnarray}
	\nu_{\mathrm{Ch}}^{j_z} = -\frac{i}{2\pi}\int_{k_{\mathrm{F}}} d^2k' \epsilon^{\alpha\beta}
    \langle \partial_{k'_{\alpha}}{u}_{n,\bm{k}}| \partial_{k'_{\beta}} {u}_{n,\bm{k}}\rangle. 
\end{eqnarray}

\subsection{Linearized gap equation}\label{sec:gapeq}
In this section, we derive the linearized gap equation for the system with time-reversal and inversion symmetries.
The mean field Hamiltonian of the superconducting system is given as 
\begin{eqnarray}
	\hat{H}(\bm{k}) = 
	\left(\begin{array}{cc}
		\bm{c}_{\bm{k}}^\dag & \bar{\bm{c}}_{-\bm{k}}^{\dag}
	\end{array}\right)
	{H}(\bm{k})
	\left(\begin{array}{c}
		\bm{c}_{\bm{k}} \\ {\bar{\bm{c}}}_{-\bm{k}}
	\end{array}\right), 
\end{eqnarray}
where the spinor $\bm{c}_{\bm{k}}$ consists of the annihilation operator $c_{j_z, \sigma_z, \bm{k}}$ 
with the indices $j_z$ and $\sigma_z$ for spin and orbital, respectively, 
its time-reversal hole partner $\bar{\bm{c}}_{-\bm{k}}$ is given as $\bar{c}_{j_z,\sigma_z,-\bm{k}}=\sum_{j_z'}(C_{2,\hat{\bm y}})_{j_zj_z'}{c}_{j_z',\sigma_z,-\bm{k}}^\dag$, 
and $H(\bm{k})$ is the BdG Hamiltonian given in Eq.~(\ref{eq:bdg_app}).
Here, we define the thermal and anomalous Green's functions
\begin{eqnarray}
	\mathcal{G}_{\bm{k}}(\tau,\tau') = -\left<T_{\tau}\left[\bm{c}_{\bm{k}}(\tau)\bm{c}_{\bm{k}}^\dag(\tau')\right]\right>, \\
	\mathcal{F}_{\bm{k}}(\tau,\tau') = -\left<T_{\tau}\left[\bar{\bm{c}}_{-\bm{k}}(\tau)\bm{c}_{\bm{k}}^\dag(\tau')\right]\right>,
\end{eqnarray}
where $\bm{c}_{\bm k}(\tau)=e^{{H_0}\tau/\hbar}\bm{c}_{\bm{k}}e^{-{H}_0\tau/\hbar}$ is the Heisenberg operator for imaginary time $\tau$. 
From the Heisenberg equation of $\bm{c}_{\bm k}(\tau)$, we obtain the Gor'kov equation of the Green's function~\cite{fetter}, 
\begin{eqnarray}
		\begin{array}{r}
		\left[i\hbar\omega_N\!-\! H_0(\bm{k})\right] \mathcal{G}_{\bm{k}}(\omega_N) \!-\! \Delta \mathcal{F}_{\bm{k}}(\omega_N)\! =\! \hbar,  \\
		\left[i\hbar\omega_N\!+\! H_0(\bm{k})\right] \mathcal{F}_{\bm{k}}(\omega_N) \!-\! \Delta \mathcal{G}_{\bm{k}}(\omega_N)\! =\! 0,
		\end{array} \label{eq:gk}
\end{eqnarray}
where we have used the time-reversal symmetry of the normal state~(\ref{eq:TRS_N}) and of the superconducting gap function~(\ref{eq:delher}).
The Fourier component of thermal and anomalous Green's functions is given by 
\begin{eqnarray}
	\mathcal{G}_{\bm{k}}(\tau,\tau') = \frac{1}{\beta\hbar}\sum_{N}e^{i\omega_{N}(\tau-\tau')}\mathcal{G}_{\bm{k}}(\omega_N),
\end{eqnarray}
with fermionic Matsubara frequency $\omega_N = (2N+1)\pi /\hbar\beta$ and $\beta= 1/k_{\mathrm{B}}T$.
In terms of the Matsubara anomalous Green's function, the gap equation is written as
\begin{eqnarray}~\label{eq:gap}
	\Delta = -\lim_{\eta\rightarrow 0}\frac{1}{\hbar\beta}\sum_{N} \int d^{3}k \tilde{V} e^{i\omega_N\eta} {\mathcal{F}}_{\bm{k}}(\omega_N), 
\end{eqnarray}
where the interaction is $\tilde{V}=U$ ($V$) when the gap function $\Delta$ is intended for intraorbital (interorbital) coupling [see also Eq.~(\ref{eq:uv})].

Near the $T_{\mathrm{c}}$ where $\Delta\rightarrow 0$, we can linearize the gap equation. 
By solving (\ref{eq:gk}) for $\mathcal{G}_{\bm{k}}$ and $\mathcal{F}_{\bm{k}}$ and 
substituting them successively, we approximate the anomalous Green's function in the lowest order of $\Delta$ as  
\begin{eqnarray}~\label{eq:Flinear}
	{\mathcal{F}}_{\bm k}(\omega_N) = -\frac{1}{\hbar} {\mathcal{G}}^0_{\bm k}(-\omega_N) \Delta {\mathcal{G}}^0_{\bm k}(\omega_N) + \mathcal{O}(\Delta^2), 
\end{eqnarray}
with the single-particle Green's function ${\mathcal{G}}^0_{\bm k}(\omega_N) = \hbar\left[ i\hbar\omega_N - H_0(\bm{k}) \right]^{-1}$. 
It is written in terms of the band representation~(\ref{eq:h0band}) as 
\begin{eqnarray}\label{eq:g0band}
	\mathcal{G}^0_{\bm k}(\omega_N) = \sum_{n} \frac{\hbar |{u}_{n,k}\rangle \langle {u}_{n,k}|}{i\hbar \omega_N - \xi_{n}(\bm{k})}.
\end{eqnarray}
By using Eq.~(\ref{eq:Flinear}) and (\ref{eq:g0band}), we can linearize the gap equation~(\ref{eq:gap}) as 
\begin{eqnarray}~\label{eq:gap2}
	\Delta &=& -\lim_{\eta\rightarrow 0}\frac{1}{\beta} \sum_{N,n,n'} \int d^{3}k \tilde{V} e^{i\omega_N\eta} \nn\\
	&&\times \frac{|u_{n,\bm{k}}\rangle \langle u_{n,\bm{k}}|\Delta |u_{n',\bm{k}}\rangle \langle u_{n',\bm{k}}|}{(-i\hbar \omega_{N}-\xi_n)(i\hbar \omega_{N}-\xi_{n'})}, 
\end{eqnarray}
In addition, the matrix element $d_{n,n'}=\langle u_{n,\bm{k}}|\Delta |u_{n',\bm{k}}\rangle$ is non-zero for ${n}'=n$ or $\bar{n}$ [see also Eq.~(\ref{eq:dweak})]
as the time-reversal or inversion partners form the Cooper pair. Hence, we reduce Eq.~(\ref{eq:gap2}) as 
\begin{eqnarray}~\label{eq:gap3}
	\Delta &&= -\lim_{\eta\rightarrow 0}\frac{1}{\beta} \sum_{N,n} \int d^{3}k \tilde{V} e^{i\omega_N\eta} \nn\\
	&&\times\!\!\! \sum_{n'=n,\bar{n}}\!\! \frac{|u_{n,\bm{k}}\rangle\langle u_{n,\bm{k}}|\Delta |u_{n',\bm{k}}\rangle\langle u_{n',\bm{k}}|}{\hbar^2 \omega_{N}^2-\xi_n^2}, \nn\\
	&&= - \sum_{n}\int d\xi_n d^{2}k' D_n(\bm{k}) \tilde{V} \frac{1}{4\xi_n}\tanh\frac{\beta\xi_n}{2}\nn\\
	&&\times\!\!\! \sum_{n'=n,\bar{n}}\!\! |u_{n,\bm{k}}\rangle\langle u_{n,\bm{k}}|\Delta |u_{n',\bm{k}}\rangle\langle u_{n',\bm{k}}|, 
\end{eqnarray}
where we have used $\xi_{\bar n}=\xi_{n}$, the expansion formula $z^{-1} \tanh(z/2)=4\sum_{n=0}^{\infty}[(2n+1)^2\pi^2+z^2]^{-1}$, 
and the momentum coordinates $(k,k_1',k_2')$ with momenta $k$ and $k'_{1,2}$ perpendicular and parallel to the Fermi surface, respectively.
$D_n(\bm{k})=\frac{d k}{d \xi_n}$ is the density of states. 

In the weak coupling limit where $\beta_c^{-1}=k_{\mathrm{B}}T_\mathrm{c}$ is much smaller than any other energy scale, 
the factor $\xi_n^{-1}\tanh(\beta_c\xi_n/2)$ has a sharp peak around the Fermi level $\xi_n=0$.
Hence, we can approximate that the other factors are represented by their values at the Fermi surface $\xi_n=0$. 
Therefore, Eq.~(\ref{eq:gap3}) is rewritten as  
\begin{eqnarray}~\label{eq:gap4}
	\Delta 
	&&= - \sum_{n}\frac{I(\beta)}{2} \int_{k_{\mathrm{F}}} d^{2}k' D_n(\bm{k}') \tilde{V} \nn\\
	&&\times\!\!\! \sum_{n'=n,\bar{n}}\!\! |u_{n,\bm{k}'}\rangle\langle u_{n,\bm{k}'}|\Delta |u_{n',\bm{k}'}\rangle\langle u_{n',\bm{k}'}|, 
\end{eqnarray}
where the energy integral is evaluated as 
$I(\beta)=\int_{-\hbar\omega_{0}}^{\hbar\omega_{0}} d\xi \frac{1}{2\xi}\tanh(\frac{\beta \xi}{2})= \ln (2e^\gamma \beta \hbar\omega_0/\pi)$ with Euler's constant $\gamma$ 
and the cutoff frequency $\omega_0$~\cite{fetter}.

In the system discussed in the main text, 
the gap function is expanded as $\Delta=\sum_{\alpha}\Delta_\alpha \Phi_{\alpha}$ [see Eq.~(\ref{eq:expand})].
By using the orthonormality of the basis $\mathrm{tr}[\Phi_{\alpha}\Phi_{\alpha'}]=\delta_{\alpha\alpha'}$ in Eq.~(\ref{eq:gap4}), 
we obtain the gap equation for $\Delta_{\alpha}$ as 
\begin{eqnarray}\label{eq:gapeq_L}
	\sum_\beta X_{\alpha\alpha'}(\beta) \Delta_{\alpha'} = 0,
\end{eqnarray}
with the coefficient
\begin{eqnarray}\label{eq:Xapp}
	\!\!\!\!\!\!
	X_{\alpha\alpha'}(\beta) &&= \delta_{\alpha\alpha'} +  
	\sum_{n} \frac{I(\beta)}{2}\!\! \int_{{k}_{\mathrm{F}}} d^2k' D_n(\bm{k}') \tilde{V} \nn\\
	&&\times \!\!\!\sum_{n'=n,\bar{n}}\!\! \langle u_{n',\bm{k}'}|\Phi_{\alpha}|u_{n,\bm{k}'}\rangle\langle u_{n,\bm{k}'}|\Phi_{\alpha'} |u_{n',\bm{k}'}\rangle.
\end{eqnarray}
The transition temperature is given by $\mathrm{det} X(\beta_c)=0$.

\begin{figure}[b]
\includegraphics[width=85mm]{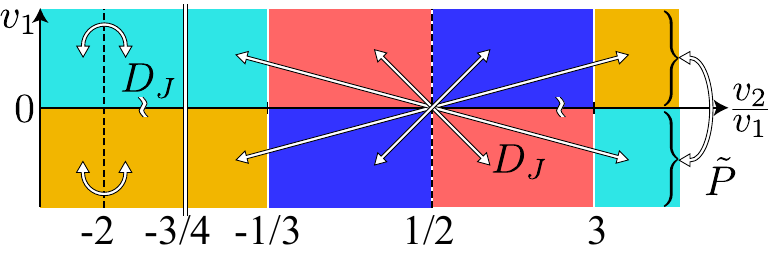}
\caption{Relations between the systems with parameters $v_1$ and $v_2/v_1$
mapped using Eq.~(\ref{eq:Dv1v2}) and (\ref{eq:Pv1v2}). 
The color code indicates the winding number $w_{\mathrm{3D}}=-2$ (turquoise), 
4 (red), $-4$ (blue), and 2 (yellow), the same as in Fig.~\ref{fig:TPdiag}.
}
\label{fig:duality}
\end{figure}

\section{Duality relation for $k\cdot p$ Hamiltonian}\label{sec:duality}
It can be easily verified that the identical vectors $\bm{J}$ and $\tilde{\bm{J}}$ in the spin space hold that
\begin{eqnarray}
	D_{J} J_i D_{J}^{\dag} = -\frac{3}{5} J_{i} - \frac{4}{5}\tilde{J}_i, \\ 
	D_{J} \tilde{J}_i D_{J}^{\dag} = -\frac{4}{5} J_{i} + \frac{3}{5}\tilde{J}_i, 
\end{eqnarray}
with the unitary matrix acting on the spin {$J=3/2$} space
\begin{eqnarray}
	D_J = \left(\begin{array}{cccc}
	 0 &0  &1 & 0 \\
	 0 &0  &0 &-1 \\
	-1 &0  &0 & 0 \\
	 0 &1  &0 & 0
	\end{array}\right).
\end{eqnarray}
As the $A_{1u}$ gap function is invariant under $D_J$, 
this unitary transformation maps the system with a set of parameters $v_1$ and $v_2$ to that with another set 
\begin{eqnarray}\label{eq:Dv1v2}
	\left(\begin{array}{c}
		v_1 \\
		v_2 
	\end{array}\right)
	\xrightarrow{D_J} 
	\left(\begin{array}{c}
		v_1' \\
		v_2' 
	\end{array}\right) = 
	\left(\begin{array}{c}
	- [3v_1 + 4 v_2]/5 \\
	- [4v_1 - 3 v_2]/5
	\end{array}\right), 
\end{eqnarray}
which yields the ``duality'' of the BdG Hamiltonian.
In addition, the operator $D_{J}$ {commutes} with the chiral operator 
\begin{eqnarray}
	\left[ \Gamma, D_J \right] = 0.
\end{eqnarray}
Therefore, the systems at the two sets of parameters in (\ref{eq:Dv1v2}) yield 
the same winding number $w_{\mathrm{3D}}$.

Similarly, the inversion operator for the superconducting state
maps the parameters as 
\begin{eqnarray} \label{eq:Pv1v2}
	\left(\begin{array}{c}
	v_1\\
	v_2
	\end{array}\right) 
	\xrightarrow{\tilde{P}}
	\left(\begin{array}{c}
	v_1'\\
	v_2'
	\end{array}\right)= 
	\left(\begin{array}{c}
	- v_1 \\
	- v_2
	\end{array}\right) 
	\hbox{ with }  \tilde{P}  = \sigma_z \tau_z.
\end{eqnarray}
Note that this mapping does not change $v_2/v_1$.
As the operator $\tilde P$ anti-commutes with the chiral operator $\Gamma$, 
\begin{eqnarray}
	\{ \Gamma, \tilde{P} \} = 0,
\end{eqnarray}
the systems at the two sets of parameters in (\ref{eq:Pv1v2}) yield 
the winding number $w_{\mathrm{3D}}$ with the opposite sign.

Finally, in Fig.~\ref{fig:duality}, 
we summarize these relations in the parameter space spanned by $v_1$ and $v_2/v_1$ used in the main text.
For $v_2/v_1>-3/4$, the unitary transformation $D_J$ maps $v_2/v_1$ to the opposite side of $v_2/v_1=1/2$ 
denoted by the dashed line in Fig.~\ref{fig:duality}. Simultaneously, it changes the sign of $v_1$. 
Furthermore, we can map {a system with $v_1$ to another with $-v_1$} using $\tilde{P}$. 
Therefore, at the opposite side with respect to $v_2/v_1=1/2$ and with the same sign as $v_1$ 
the winding number $w_{\mathrm{3D}}$ exhibits the opposite sign. 
In contrast, for $v_2/v_1<-3/4$, the unitary transformation $D_J$ {connects} the 
parameters with the same sign as $v_1$ (see Fig.~\ref{fig:duality}), 
which have the {same} winding number $w_{\mathrm{3D}}$. 

\bibliography{reference}
\end{document}